\documentclass[11pt]{article}
\usepackage[a4paper, left=2.5cm, right=2.5cm, top=2.5cm, bottom=2.5cm]{geometry}
\usepackage{graphicx} 
\usepackage{caption}
\usepackage{subcaption} 
\usepackage{hyperref} 
\usepackage{enumitem}
\usepackage[round, authoryear]{natbib}
\usepackage{amsmath, amssymb, bm}
\usepackage{amsfonts}
\usepackage{comment}
\usepackage{mathtools} 
\mathtoolsset{showonlyrefs} 
\usepackage{booktabs} 

\usepackage{authblk}

\title{Memory-Induced Supra-Competitive Outcomes Between Deep Reinforcement Learning Agents in Optimal Trade Execution}

\date{\today} 

\author[1]{Christos S. Koulouris}
\author[1,2]{Carlo Campajola}

\affil[1]{Institute of Finance and Technology, University College London, Gower Street WC1E 6BT London, United Kingdom}
\affil[2]{UZH Blockchain Center, Andreasstrasse 15, 8050 Z\"{u}rich, Switzerland}

\usepackage{upgreek}
\usepackage{algorithm}
\usepackage{algorithmic}
\usepackage{subfiles}

\usepackage{amsthm}
\theoremstyle{remark}

\begin{document}

\maketitle

\begin{abstract}
In this paper, we investigate whether deep reinforcement-learning agents interacting in a shared optimal-execution environment can sustain supra-competitive outcomes, in the sense of achieving lower implementation shortfalls than the relevant game-theoretical competitive benchmark. We study a two-agent Almgren--Chriss liquidation game and examine how learned behavior depends on intra-episode environment feedback, the ability to interpret the mid-price and the agent's knoledge of the past. We first use ex-ante schedule-learning agents to remove intra-episode feedback and isolate what can arise when agents commit to complete liquidation trajectories before execution begins. We then allow agents to condition on the evolving state using a variety of DDQN architectures. We find that, when agents are given access to intra-episode history, especially recent prices and own past actions, supra-competitive outcomes become substantially more frequent and more persistent. These findings indicate that supra-competitive behavior in this execution game is driven not by multi-agent learning or by current price observation alone, but by feedback, memory, and state-contingent interaction along the realized execution path.
\end{abstract}

\section{Introduction}

Optimal trade execution is a canonical problem in market microstructure. In the classical formulation, a trader splits a large parent order over a finite horizon so as to balance market impact against price risk \citep{almgren2001optimal}. While this problem has traditionally been studied through stochastic control under stylized assumptions, deep reinforcement learning (DRL) has opened a more flexible route that can accommodate richer state representations, more complex dynamics, and less restrictive policy classes \citep{gould2013limit,hambly2023recent}. This has led to a growing literature on DRL-based execution, ranging from early reinforcement-learning formulations of execution scheduling to more recent deep value-based and actor--critic approaches \citep{nevmyvaka2006reinforcement,ning2018double,daberius2019deep,moallemi2022reinforcement,lin2020endtoend,byun2023practical,macri2025reinforcement}. Recent work has also expanded the execution problem beyond pure liquidation-rate control to richer limit-order-book environments in which agents allocate dynamically between market and limit orders \citep{cheridito2026marketlimit}.

Most of this literature is single-agent. Yet execution is inherently strategic once multiple large traders interact through shared price impact. In that case, what appears as a control problem from the viewpoint of one trader becomes a game, and the natural theoretical benchmark is no longer a single-agent optimum but a Nash equilibrium. This connects optimal execution to a broader literature on repeated games with learning agents, where supra-competitive outcomes have been documented in a range of settings. A recurring message from that literature is that such outcomes are not driven by a single mechanism. Rather, they have been linked to reward--punishment dynamics, limited or decaying exploration, coarse action spaces, and more generally to feedback structures that let agents condition on one another's past behavior \citep{calvano2020artificial,klein2021autonomous,deng2025exploring,xu2024mechanism,denboer2022artificial,carissimo2025orchestration}. In market microstructure environments, related phenomena have also been reported in dealer markets, market-making games, and execution games, where agents interact through shared liquidity, inventory pressure, and endogenous price formation rather than through posted prices alone \citep{cont2024dealer,han2022cooperation,lillo2024deviations,cartea2022tick,dou2025aipowered,foscari2025invisible}.

Our paper continues the line of work initiated by \citet{lillo2024deviations}, who study two deep RL agents in an Almgren--Chriss execution game and show that learned behavior can move away from the Nash benchmark and toward the Pareto frontier. We revisit this question and focus on the mechanism behind such deviations. In particular, we ask whether supra-competitive outcomes are a generic consequence of multi-agent learning in this execution game, or whether they depend on the information available to the agents during execution: current price observations, intra-episode feedback, and the realized history of prices and actions.

We make three main contributions:
\begin{itemize}
    \item We introduce ex-ante schedule-learning agents that choose complete liquidation trajectories before the episode begins, which, to the best of our knowledge has not been done in previous optimal trade execution literature. This removes within-episode feedback, and therefore removes channels for implicit signalling, punishment-like responses, and history-dependent adaptation. By examining various environment design choices, we suggest that limited information flow can make the learners behave as if a different effective benchmark is relevant, leading sometimes to situations that could be interpreted as supra-competitive.

    \item We show that allowing agents to condition on the evolving state is not, by itself, sufficient to generate stable supra-competitive outcomes. We implement a simultaneous-move DDQN environment, similar to previous literature. In this setting, baseline DDQN agents remain close to the competitive benchmark. Moreover, architectures that make the price coordinate more salient increase price attribution but still do not produce persistent supra-competitive behavior.

    \item We show that intra-episode history is the key channel through which stable supra-competitive outcomes arise. When agents are allowed to condition on recent prices and their own past actions, outcomes below the Nash benchmark become substantially more frequent and more persistent, while remaining bounded away from the Pareto-efficient TWAP benchmark.
\end{itemize}

The remainder of the paper is structured as follows. Section~\ref{sec:optimal-execution-environment} presents the multi-agent Almgren--Chriss execution environments and the benchmark schedules used for comparison. Section~\ref{sec:open_loop_rl} studies ex-ante schedule learning, where intra-episode feedback is removed. Section~\ref{sec:baseline-experiment} introduces the baseline DDQN experiment, in which agents can condition on the evolving state. The following sections examine whether stronger price conditioning and explicit intra-episode history are sufficient to generate persistent supra-competitive outcomes. Section~\ref{sec:conclusions} concludes.
\section{The Almgren--Chriss Trade Execution Game}
\label{sec:optimal-execution-environment}

To the best of our knowledge, \citet{lillo2024deviations} provide the first dedicated reinforcement-learning study documenting supra-competitive outcomes in a multi-agent optimal-execution game, and we therefore take that paper as our closest point of departure. Both their study and ours train learning agents in discrete Almgren--Chriss execution environments. However, the relevant competitive benchmark depends on a modelling choice that is easy to overlook: whether temporary impact is computed from aggregate same-period order flow or from each agent's own order only.

This distinction is central for the analysis below. In the aggregate-temporary-impact environment, the natural theoretical reference is the open-loop Nash equilibrium of \citet{schied2017game}, whose affected price contains both aggregate permanent impact and aggregate temporary impact. In the own-temporary-impact environment, the relevant benchmark is the discrete equilibrium of \citet{cordoni2024transient}, specialized to the constant-kernel case corresponding to permanent impact. Alongside these competitive benchmarks, we also report the TWAP schedule. This is not only a simple execution baseline: in the symmetric risk-neutral two-player setting, \citet{lillo2024deviations} show that TWAP coincides with the Pareto-efficient liquidation schedule. It therefore provides the natural cooperative benchmark against which deviations from Nash can be interpreted.

Throughout the paper, the reinforcement-learning agents trade on a finite grid. Therefore, although the continuous-time formulas are useful for identifying the underlying game and for relating our setting to \citet{schied2017game}, the benchmark points used in the empirical comparisons are the discrete-time, grid-implemented Nash equilibria. This ensures that learned policies are compared with benchmark strategies that are actually implementable in the same simulator. In the sections that follow, learned outcomes are therefore classified relative to the discrete Nash benchmark corresponding to the temporary-impact convention of the environment in which the agents are trained, and their distance from TWAP is used to assess how close they are to the Pareto-efficient cooperative benchmark.

\subsection{Aggregate temporary impact}
\label{subsec:aggregate-temp-impact}

Consider two agents who must liquidate inventories \(q_{1,0}>0\) and \(q_{2,0}>0\) over a fixed horizon \([0,T]\). Let \(q_i(t)\) denote the remaining inventory of agent \(i\), with
\[
q_i(0)=q_{i,0},
\qquad
q_i(T)=0,
\qquad i\in\{1,2\},
\]
and let
\[
u_i(t)\coloneqq -\dot q_i(t)
\]
be the liquidation rate. The unaffected price follows
\[
S_t^0=S_0+\sigma W_t,
\]
where \(W\) is a standard Brownian motion and \(\sigma>0\) is the exogenous volatility parameter.

In the aggregate-temporary-impact formulation, the transaction price is
\begin{equation}
\widetilde S^{\mathrm{agg}}(t)
=
S_0
-\kappa \int_0^t \sum_{j=1}^2 u_j(s)\,ds
-a \sum_{j=1}^2 u_j(t)
+\sigma W_t,
\label{eq:ct_agg_exec_price}
\end{equation}
where \(\kappa>0\) is the permanent-impact coefficient and \(a>0\) is the temporary-impact coefficient. The important feature of \eqref{eq:ct_agg_exec_price} is that both permanent and temporary impact are generated by aggregate trading:
\[
-\kappa \int_0^t \sum_{j=1}^2 u_j(s)\,ds,
\qquad
-a\sum_{j=1}^2 u_j(t).
\]
This mirrors the continuous-time aggregate-impact formulation of \citet{schied2017game} (equation~(6) in that article), in which the affected price contains the temporary-impact term
\[
\lambda\sum_{j=1}^n \dot X_j(t).
\]
With our sign convention, liquidation rates are positive, so \(u_j(t)=-\dot X_j(t)\). Hence, the Schied--Zhang aggregate temporary-impact term is represented in our notation by the total liquidation rate \(\sum_j u_j(t)\). Therefore, the Schied--Zhang transaction price corresponds to aggregate temporary impact, not own temporary impact.

For a deterministic liquidation strategy \(u_i\), given the opponent's strategy \(u_{-i}\), the implementation shortfall of agent \(i\) is
\begin{equation}
\mathrm{IS}_i(u_i\mid u_{-i})
\coloneqq
q_{i,0}S_0
-
\int_0^T u_i(t)\,\widetilde S^{\mathrm{agg}}(t)\,dt.
\label{eq:ct_agg_is_def}
\end{equation}
The mean--variance objective is
\begin{equation}
J_i(u_i\mid u_{-i})
\coloneqq
\mathbb E\!\left[\mathrm{IS}_i(u_i\mid u_{-i})\right]
+\frac{\lambda}{2}\,
\mathrm{Var}\!\left[\mathrm{IS}_i(u_i\mid u_{-i})\right],
\label{eq:ct_agg_objective}
\end{equation}
where \(\lambda\ge 0\) is the risk-aversion parameter. Using \eqref{eq:ct_agg_exec_price}--\eqref{eq:ct_agg_is_def}, the expected shortfall is
\begin{equation}
\mathbb E\!\left[\mathrm{IS}_i(u_i\mid u_{-i})\right]
=
\kappa
\int_0^T
u_i(t)
\left(
\int_0^t \sum_{j=1}^2 u_j(s)\,ds
\right)dt
+
a
\int_0^T
u_i(t)\sum_{j=1}^2 u_j(t)\,dt.
\label{eq:ct_agg_expected_is}
\end{equation}
The first term captures the cumulative permanent price pressure generated by aggregate liquidation, while the second term captures contemporaneous aggregate temporary impact.

For the two-player aggregate-impact game, \citet{schied2017game} show that the unique open-loop Nash equilibrium admits a closed-form representation in terms of the inventory paths
\begin{equation}
q_1^*(t)=\frac{1}{2}\bigl(\Sigma(t)+\Delta(t)\bigr),
\qquad
q_2^*(t)=\frac{1}{2}\bigl(\Sigma(t)-\Delta(t)\bigr),
\label{eq:sz_qstar_general}
\end{equation}
where
\begin{equation}
\Sigma(t)
=
Q\,
\exp\!\left(-\frac{\kappa t}{6a}\right)
\frac{
\sinh\!\bigl((T-t)\nu_{\Sigma}\bigr)
}{
\sinh\!\bigl(T\nu_{\Sigma}\bigr)
},
\qquad
\nu_{\Sigma}
\coloneqq
\frac{\sqrt{\kappa^2+12a\lambda\sigma^2}}{6a},
\label{eq:sz_sigma_general}
\end{equation}
\begin{equation}
\Delta(t)
=
\widetilde Q\,
\exp\!\left(\frac{\kappa t}{2a}\right)
\frac{
\sinh\!\bigl((T-t)\nu_{\Delta}\bigr)
}{
\sinh\!\bigl(T\nu_{\Delta}\bigr)
},
\qquad
\nu_{\Delta}
\coloneqq
\frac{\sqrt{\kappa^2+4a\lambda\sigma^2}}{2a},
\label{eq:sz_delta_general}
\end{equation}
with
\[
Q\coloneqq q_{1,0}+q_{2,0},
\qquad
\widetilde Q\coloneqq q_{1,0}-q_{2,0}.
\]
The corresponding Nash liquidation rates are \(u_i^*(t)=-\dot q_i^*(t)\).

The experiments below focus on the symmetric risk-neutral case,
\[
q_{1,0}=q_{2,0}=q_0,
\qquad
\lambda=0.
\]
Then \(\widetilde Q=0\), the two players use the same inventory path, and \eqref{eq:sz_qstar_general}--\eqref{eq:sz_delta_general} simplify to
\begin{equation}
q^{\mathrm{N,agg}}(t)
=
q_0
\frac{
\exp\!\bigl(\beta_{\mathrm{agg}}(T-t)\bigr)-1
}{
\exp(\beta_{\mathrm{agg}}T)-1
},
\qquad
\beta_{\mathrm{agg}}\coloneqq \frac{\kappa}{3a}.
\label{eq:agg_ct_inventory}
\end{equation}
The liquidation rate is
\begin{equation}
u^{\mathrm{N,agg}}(t)
=
-\frac{d}{dt}q^{\mathrm{N,agg}}(t)
=
\beta_{\mathrm{agg}} q_0
\frac{
\exp\!\bigl(\beta_{\mathrm{agg}}(T-t)\bigr)
}{
\exp(\beta_{\mathrm{agg}}T)-1
}.
\label{eq:agg_ct_rate}
\end{equation}
Substituting \eqref{eq:agg_ct_rate} into \eqref{eq:ct_agg_expected_is} gives the analytical continuous-time aggregate-impact Nash cost,
\begin{equation}
\mathrm{IS}_i^{\mathrm{N,agg,ct}}
=
\kappa q_0^2
\left[
1+\frac{1}{3}\coth\!\left(\frac{\kappa T}{6a}\right)
\right].
\label{eq:agg_ct_nash_is}
\end{equation}
This analytical quantity is useful as the continuous-time reference for the Schied--Zhang game. However, the agents in our experiments trade on a finite grid, and therefore the operational benchmark is the grid-implemented version of the inventory path \eqref{eq:agg_ct_inventory}.

Let
\[
t_m\coloneqq m\tau,
\qquad
m=0,1,\dots,N,
\qquad
\tau\coloneqq \frac{T}{N}.
\]
The permanently impacted, discrete-time price process is given by
\begin{equation}
S_t
=
S_{t-1}
-\kappa V_t
+\sigma\sqrt{\tau}\,\xi_t,
\qquad
\xi_t\sim\mathcal N(0,1),
\label{eq:disc_agg_mid}
\end{equation}
and the inventory held by agent $k$ at each timestep is
\[
q_t^{(k)}=q_{t-1}^{(k)}-v_t^{(k)},
\qquad
V_t=\sum_{k=1}^K v_t^{(k)}.
\]
The common execution price, taking temporary impact from both traders at once, is then
\begin{equation}
\widetilde S_t^{\mathrm{agg}}
=
S_{t-1}
-a\,\frac{V_t}{\tau}.
\label{eq:disc_agg_exec}
\end{equation}
Agent $k$'s cash process evolves as
\begin{equation}
C_t^{(k)}
=
C_{t-1}^{(k)}
+
v_t^{(k)}\widetilde S_t^{\mathrm{agg}},
\label{eq:disc_agg_cash}
\end{equation}
and complete liquidation is enforced at the terminal date:
\[
q_N^{(k)}=0.
\]

The grid-implemented aggregate Nash schedule is obtained by forward differences of \eqref{eq:agg_ct_inventory}:
\begin{equation}
v_m^{\mathrm{N,agg},\Delta}
\coloneqq
q^{\mathrm{N,agg}}(t_{m-1})
-
q^{\mathrm{N,agg}}(t_m),
\qquad
m=1,\dots,N.
\label{eq:agg_forward_difference}
\end{equation}
Equivalently,
\begin{equation}
v_m^{\mathrm{N,agg},\Delta}
=
q_0\,
\frac{
\exp\!\bigl(\beta_{\mathrm{agg}}(T-t_m)\bigr)
\bigl(\exp(\beta_{\mathrm{agg}}\tau)-1\bigr)
}{
\exp(\beta_{\mathrm{agg}}T)-1
},
\qquad
\beta_{\mathrm{agg}}=\frac{\kappa}{3a}.
\label{eq:disc_agg_nash_actions}
\end{equation}
The corresponding discrete implementation shortfall is computed inside the aggregate temporary-impact simulator:
\begin{equation}
\mathrm{IS}_i^{\mathrm{N,agg},\Delta}
=
q_0S_0
-
\sum_{m=1}^N
v_m^{\mathrm{N,agg},\Delta}
\widetilde S_m^{\mathrm{N,agg},\Delta},
\label{eq:disc_agg_nash_cost}
\end{equation}
where \(\widetilde S_m^{\mathrm{N,agg},\Delta}\) is generated recursively from \eqref{eq:disc_agg_mid}--\eqref{eq:disc_agg_exec} using
\[
V_m^{\mathrm{N,agg},\Delta}
=
2v_m^{\mathrm{N,agg},\Delta}
\]
in the symmetric two-player case.

As mentioned, the natural benchmark for supra-competitive outcomes is the Pareto-optimal TWAP schedule
\begin{equation}
v_m^{\mathrm{TWAP},\Delta}
=
\frac{q_0}{N},
\qquad
m=1,\dots,N,
\label{eq:disc_twap_actions_agg}
\end{equation}
and under the aggregate temporary impact environment, its implementation shortfall is
\begin{equation}
\mathrm{IS}_i^{\mathrm{TWAP,agg},\Delta}
=
\kappa q_0^2\frac{N-1}{N}
+
\frac{2a q_0^2}{T}.
\label{eq:disc_twap_agg_cost}
\end{equation}
The factor \(2a\) appears because each player faces temporary impact generated by total symmetric slice flow \(V_t=2q_0/N\).

\subsection{Own temporary impact}
\label{subsec:own-temp-impact}

We also consider an own-temporary-impact version of the execution game. This is the convention closest to the discrete simulator used by \citet{lillo2024deviations}: the midprice is affected by aggregate order flow, but the temporary-impact component of the execution price depends only on the agent's own trade. Thus, relative to the aggregate-temporary-impact environment of Section~\ref{subsec:aggregate-temp-impact}, the permanent-impact channel remains strategic and common, while the temporary-impact channel becomes individual.

On the finite trading grid, the midprice evolves as
\begin{equation}
S_t
=
S_{t-1}
-\kappa V_t
+\sigma\sqrt{\tau}\,\xi_t,
\qquad
V_t=\sum_{k=1}^K v_t^{(k)},
\qquad
\xi_t\sim\mathcal N(0,1),
\label{eq:disc_own_mid}
\end{equation}
where \(v_t^{(k)}\) is the trade of agent \(k\) at slice \(t\), \(V_t\) is total slice flow, and \(\tau=T/N\) is the grid spacing. Inventories satisfy
\[
q_t^{(k)}=q_{t-1}^{(k)}-v_t^{(k)},
\qquad
q_N^{(k)}=0.
\]
The difference from the aggregate-temporary-impact environment is the execution price. Under own temporary impact, agent \(k\) receives
\begin{equation}
\widetilde S_t^{(k),\mathrm{own}}
=
S_{t-1}
-
a\,\frac{v_t^{(k)}}{\tau},
\label{eq:disc_own_exec}
\end{equation}
and cash evolves as
\begin{equation}
C_t^{(k)}
=
C_{t-1}^{(k)}
+
v_t^{(k)}\widetilde S_t^{(k),\mathrm{own}}.
\label{eq:disc_own_cash}
\end{equation}
Thus, the other agent affects agent \(k\) through future midprices via \eqref{eq:disc_own_mid}, but not through the same-slice temporary-impact term in \eqref{eq:disc_own_exec}.

This convention can be motivated in two equivalent ways. First, it corresponds to a modification of the Schied--Zhang aggregate-impact game in which the temporary-impact term is made own-order. In continuous time, this changes the expected implementation shortfall from the aggregate form
\[
\kappa
\int_0^T
u_i(t)
\left(
\int_0^t \sum_{j=1}^2 u_j(s)\,ds
\right)dt
+
a
\int_0^T
u_i(t)\sum_{j=1}^2 u_j(t)\,dt
\]
to
\begin{equation}
\mathbb E\!\left[\mathrm{IS}_i^{\mathrm{own}}(u_i\mid u_{-i})\right]
=
\kappa
\int_0^T
u_i(t)
\left(
\int_0^t \sum_{j=1}^2 u_j(s)\,ds
\right)dt
+
a
\int_0^T
u_i(t)^2\,dt.
\label{eq:ct_own_expected_is}
\end{equation}
The permanent-impact term is unchanged, because permanent price pressure is still generated by aggregate liquidation. The temporary-impact term changes from \(a u_i(t)\sum_j u_j(t)\) to \(a u_i(t)^2\), because agent \(i\)'s temporary execution cost depends only on its own rate. In the symmetric risk-neutral two-player case, the corresponding Nash inventory path has the same exponential form as the aggregate case,
\begin{equation}
q^{\mathrm{N,own,ct}}(t)
=
q_0
\frac{
\exp\!\bigl(\beta_{\mathrm{own}}(T-t)\bigr)-1
}{
\exp(\beta_{\mathrm{own}}T)-1
},
\qquad
\beta_{\mathrm{own}}
=
\frac{\kappa}{2a}.
\label{eq:own_ct_inventory}
\end{equation}
Since \(\kappa/(2a)>\kappa/(3a)\), the own-temporary-impact Nash path is more front-loaded than the Schied--Zhang aggregate-temporary-impact Nash path.

Second, and more importantly for the discrete experiments, the own-temporary-impact benchmark can be written directly in discrete time. We use the constant-kernel Nash equilibrium of \citet{cordoni2024transient}, specialized to the case in which the transient-impact kernel is constant. In the present Almgren--Chriss setting, this constant-kernel case corresponds to permanent impact. Rewritten in our notation, the relevant permanent-impact coefficient is \(\kappa\), while the temporary cost generated by the own-temporary-impact execution price in \eqref{eq:disc_own_exec} is
\[
a\,\frac{\left(v_t^{(k)}\right)^2}{\tau}.
\]
Thus, the discrete own-temporary-impact benchmark is characterized by the permanent-impact coefficient \(\kappa\) and the own temporary-cost coefficient \(a/\tau\).

The general two-player equilibrium in \citet{cordoni2024transient} can be decomposed into a symmetric component and an antisymmetric component. Since our experiments use symmetric initial inventories,
\[
q_{1,0}=q_{2,0}=q_0,
\]
the antisymmetric component vanishes. Therefore both agents use the same normalized trade profile. In the constant-kernel case, this profile is geometric. For \(N\) trading slices, define
\begin{equation}
\Lambda
=
\frac{2a}{\kappa\tau}
+
\frac{1}{2},
\qquad
\rho
=
1-\frac{1}{\Lambda}.
\label{eq:own_discrete_rho}
\end{equation}
Then the own-temporary-impact Nash trade of each agent at slice \(m\) is
\begin{equation}
v_m^{\mathrm{N,own},\Delta}
=
q_0
\frac{(1-\rho)\rho^{m-1}}{1-\rho^N},
\qquad
m=1,\ldots,N.
\label{eq:disc_own_nash_actions}
\end{equation}
The normalization ensures that
\[
\sum_{m=1}^N v_m^{\mathrm{N,own},\Delta}=q_0,
\]
so the terminal liquidation constraint is satisfied exactly. The auxiliary variable \(\rho\) is the geometric decay ratio implied by the discrete constant-kernel equilibrium.

The associated implementation shortfall is evaluated directly in the own-temporary-impact simulator:
\begin{equation}
\mathrm{IS}_i^{\mathrm{N,own},\Delta}
=
q_0S_0
-
\sum_{m=1}^N
v_m^{\mathrm{N,own},\Delta}
\widetilde S_m^{(i),\mathrm{N,own},\Delta},
\label{eq:disc_own_nash_cost}
\end{equation}
where \(\widetilde S_m^{(i),\mathrm{N,own},\Delta}\) is generated recursively from \eqref{eq:disc_own_mid}--\eqref{eq:disc_own_exec}. In the symmetric case,
\[
v_m^{(1),\mathrm{N,own},\Delta}
=
v_m^{(2),\mathrm{N,own},\Delta}
=
v_m^{\mathrm{N,own},\Delta},
\qquad
V_m^{\mathrm{N,own},\Delta}
=
2v_m^{\mathrm{N,own},\Delta}.
\]

The continuous-time modification in \eqref{eq:own_ct_inventory} is useful for intuition. It implies a geometric decay ratio approximately equal to
\[
\exp\!\left(-\frac{\kappa\tau}{2a}\right)
\]
on a fine grid. The discrete constant-kernel benchmark above gives the exact finite-grid ratio \(\rho\) in \eqref{eq:own_discrete_rho}. In the experiments below, we therefore use \eqref{eq:disc_own_nash_actions}, not the continuous-time approximation, as the own-temporary-impact Nash benchmark.

The TWAP schedule is the same inventory path as in the aggregate-temporary-impact environment:
\begin{equation}
v_m^{\mathrm{TWAP},\Delta}
=
\frac{q_0}{N},
\qquad
m=1,\ldots,N.
\label{eq:disc_twap_actions_own}
\end{equation}
However, its implementation shortfall differs because the execution price is different. Under own temporary impact,
\begin{equation}
\mathrm{IS}_i^{\mathrm{TWAP,own},\Delta}
=
\kappa q_0^2\frac{N-1}{N}
+
\frac{a q_0^2}{T}.
\label{eq:disc_twap_own_cost}
\end{equation}
The permanent-impact term is the same as in the aggregate-temporary-impact environment, because the midprice is still driven by aggregate flow. The temporary-impact term is smaller because each player pays temporary impact only on its own slice flow \(q_0/N\), rather than on the total symmetric slice flow \(2q_0/N\).

In the empirical sections that follow, all own-temporary-impact experiments are evaluated relative to the discrete Nash benchmark \eqref{eq:disc_own_nash_actions}. This is the benchmark corresponding to the own-temporary-impact environment in which the agents are trained. It should therefore be distinguished from the Schied--Zhang aggregate-temporary-impact benchmark used for the aggregate environment.

\begin{figure}[t]
    \centering
    \includegraphics[width=0.95\textwidth]{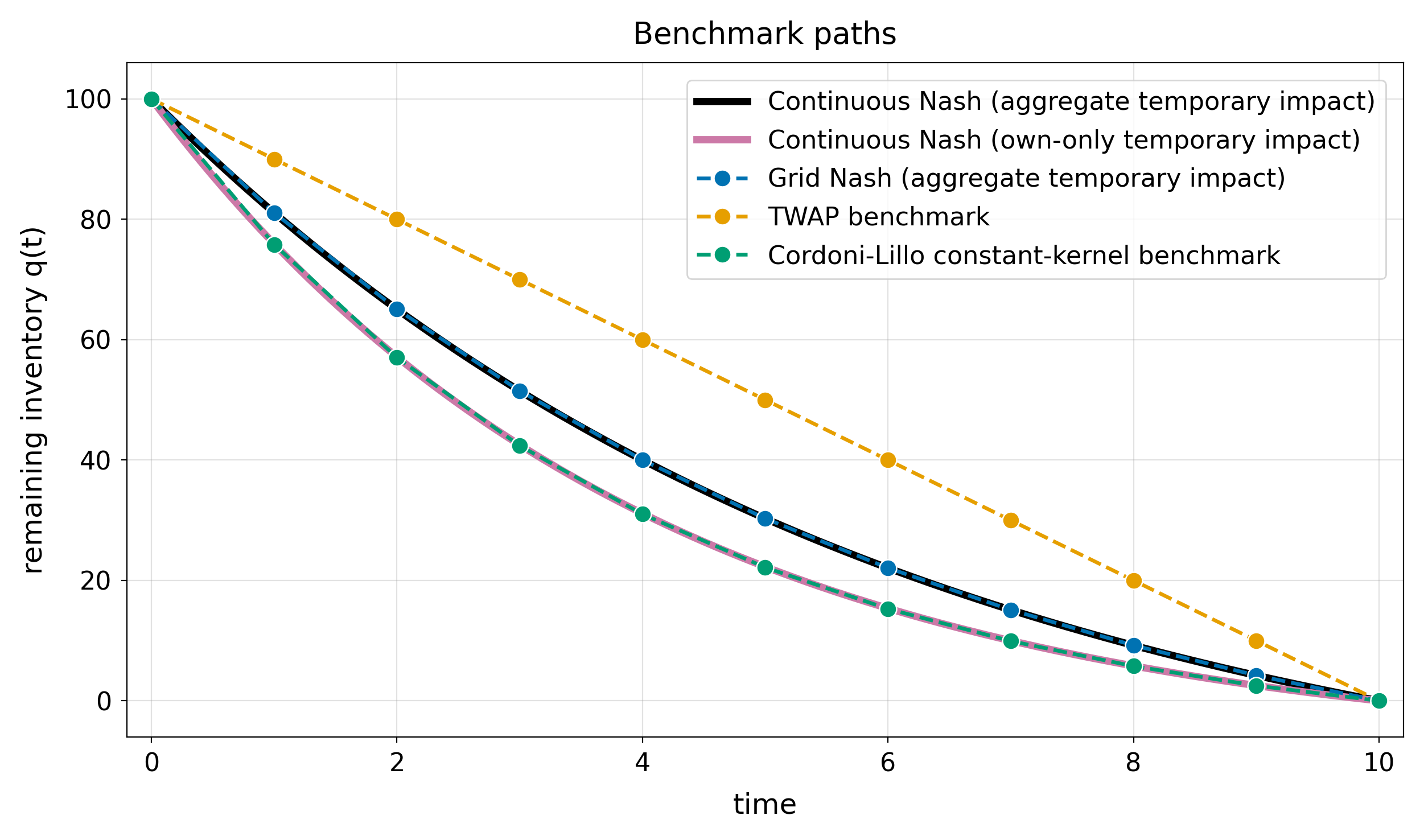}
    \caption{
    Benchmark inventory paths for the symmetric risk-neutral two-player setting.
    The aggregate-temporary-impact Nash path is obtained by grid-implementing the
    Schied--Zhang equilibrium with \(\beta_{\mathrm{agg}}=\kappa/(3a)\).
    The own-temporary-impact benchmark is obtained from the Cordoni--Lillo
    constant-kernel discrete equilibrium. The TWAP path is included as the
    Pareto-efficient cooperative reference.
    }
    \label{fig:benchmark_paths}
\end{figure}
\section{Ex-ante Schedule Learning}
\label{sec:open_loop_rl}

The competitive benchmarks presented in Section~\ref{sec:optimal-execution-environment} are equilibrium schedules: each agent chooses a full liquidation trajectory over the trading horizon, rather than a state-dependent rule that is revised after observing the realized intra-episode price path. In this section, we therefore introduce schedule-based deep reinforcement-learning agents that are forced to make the same type of commitment. Each agent chooses an entire action sequence at the beginning of the episode and then executes it without updating its trades in response to within-episode price movements, realized inventories, or the opponent's realized actions.

This experiment is useful because it removes intra-episode feedback. In a state-dependent learning problem, an agent can observe information generated during the episode---such as price movements, remaining inventory, previous aggregate flow, or other traces of the interaction---and use that information to adjust later actions. Such feedback can create channels for implicit signalling and punishment-like responses, mechanisms that are often associated with supra-competitive outcomes in learning environments. By contrast, the schedule-based agents studied here cannot react within an episode. Once chosen, the schedule does not change throughout the episode. They therefore provide a control experiment: if deviations from Nash appear even after removing intra-episode feedback, then those deviations cannot be attributed solely to within-episode state dependence or history-based interaction.

We study an ex-ante schedule-learning procedure inspired by the Pontryagin approach of \citet{eberhard2025pontryagin}. Below we focus on the model-free version, where schedule updates are estimated from perturbation rollouts of the execution environment. A model-based version gives qualitatively similar results and is described in~\ref{app:deterministic_surrogate_schedule_learning}.

\subsection{Model-free schedule learning}
\label{subsec:mf_open_loop}

For the model-free schedule learner, player \(k\in\{1,2\}\) is represented by a time-only neural network. The network
\[
\phi_{\theta^{(k)}}:[0,1]\to\mathbb R
\]
takes normalized time \(t/N\) as input and produces a scalar score for each trading slice. These scores are combined with a learnable slice-specific bias \(b_t^{(k)}\), so that
\[
z_t^{(k)}
=
\phi_{\theta^{(k)}}(t/N)+b_t^{(k)},
\qquad
t=0,\dots,N-1.
\]
Here, \(\phi_{\theta^{(k)}}\) is a neural network. Its role is to learn the smooth time profile of the liquidation schedule, while the bias terms allow additional flexibility at individual trading dates.

The raw scores are converted into non-negative trading weights and normalized to satisfy the liquidation constraint exactly:
\begin{equation}
w_t^{(k)}
=
\mathrm{softplus}\!\bigl(z_t^{(k)}\bigr),
\qquad
U_t^{(k)}
=
q_0^{(k)}
\frac{w_t^{(k)}}{\sum_{s=0}^{N-1}w_s^{(k)}}.
\label{eq:ol_mf_schedule}
\end{equation}
Hence
\[
U_t^{(k)}\ge 0,
\qquad
\sum_{t=0}^{N-1}U_t^{(k)}=q_0^{(k)}.
\]
The learned object is therefore a complete liquidation schedule
\[
\bm U_\theta^{(k)}
=
\bigl(U_0^{(k)},\ldots,U_{N-1}^{(k)}\bigr),
\]
chosen before the episode begins. Once this schedule is fixed, the agent executes it without conditioning on the realized intra-episode price path, remaining inventory, or opponent behavior.

Training is based on simulator rollouts. At each update, the current pair of schedules is executed in the environment, and each agent estimates how small perturbations of its own schedule affect its cumulative reward, while the opponent's schedule is held fixed. Perturbations are constructed to preserve the liquidation constraint, so that the perturbed schedule still sums to the initial inventory. Using common random numbers across the baseline and perturbed rollouts reduces the variance of this comparison.

Let
\[
R^{(k)}(\bm U^{(k)},\bm U^{(-k)})
=
\sum_{t=0}^{N-1} r_t^{(k)}
\]
denote the cumulative reward returned by the execution environment to player \(k\). For perturbations \(\epsilon_\ell^{(k)}\) of player \(k\)'s schedule, the local schedule-gradient direction \(g^{(k)}\) is estimated from
\begin{equation}
R^{(k)}\!\left(\bm U^{(k)}+\epsilon_\ell^{(k)},\bm U^{(-k)}\right)
-
R^{(k)}\!\left(\bm U^{(k)},\bm U^{(-k)}\right)
\approx
g^{(k)\top}\epsilon_\ell^{(k)}.
\label{eq:ol_mf_reward_gradient_fit}
\end{equation}
In practice, \(g^{(k)}\) is obtained by a ridge-regression fit over multiple perturbation rollouts. The policy parameters are then updated by ascending the first-order surrogate objective
\begin{equation}
\mathcal L_{\mathrm{EA}}^{(k)}(\theta^{(k)})
=
-\frac{1}{N}
g^{(k)\top}
\bm U_\theta^{(k)}.
\label{eq:ol_mf_loss}
\end{equation}
The negative sign appears because the implementation minimizes this loss using gradient descent, which corresponds to moving the schedule in the estimated reward-improving direction.

The important implementation point is that rewards are supplied by the execution environment itself. Therefore, the temporary-impact convention enters through the simulator: when the agent is trained in the aggregate-temporary-impact environment, rewards are generated using the aggregate execution price, while in the own-temporary-impact environment they are generated using the own execution price. Experimental performance is always assessed using the implementation shortfall computed from the simulator's realized cash process.

\subsection{Results from schedule learning}
\label{subsec:open_loop_results}

\begin{figure}[!htbp]
    \centering
    \includegraphics[width=0.95\textwidth]{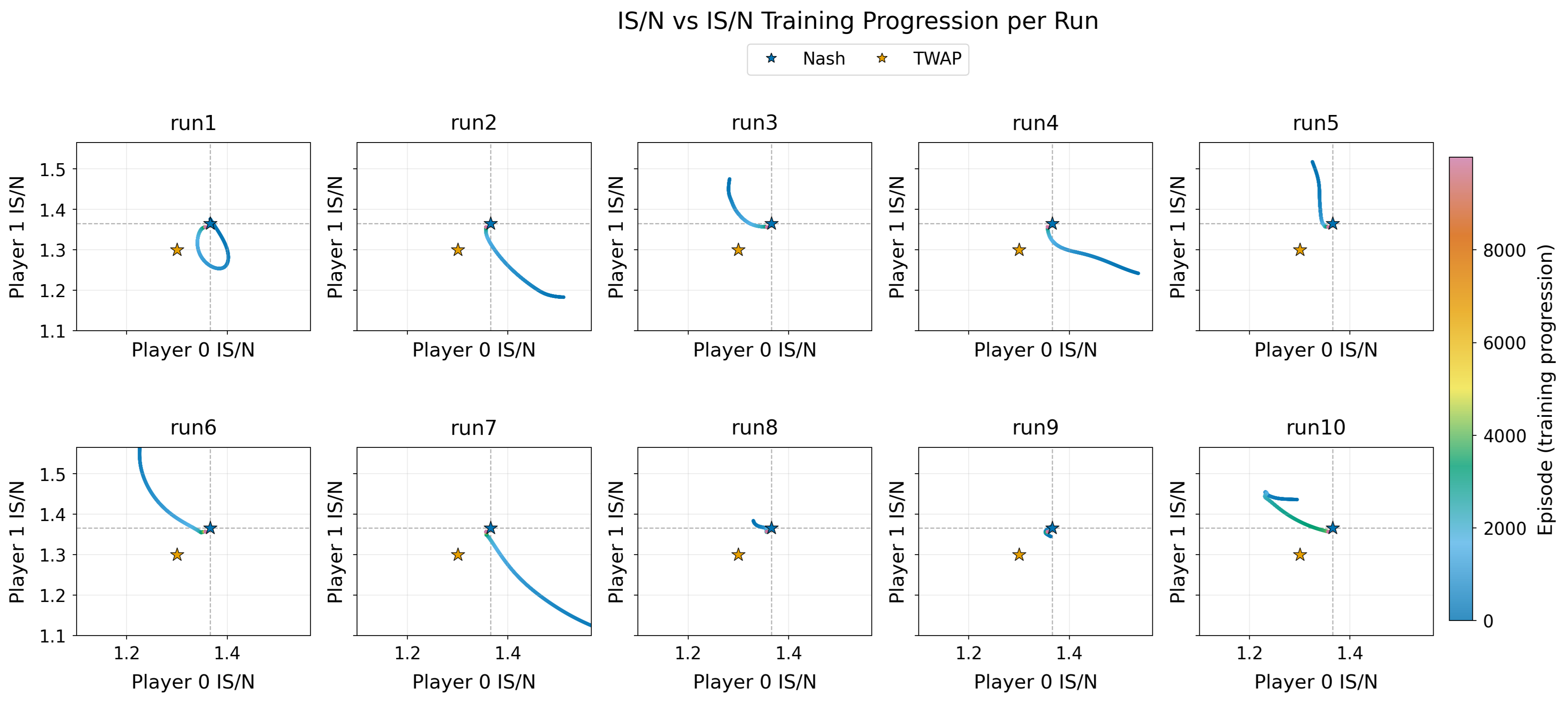}
    \caption{Schedule-learning results in the aggregate-temporary-impact environment.}
    \label{fig:open_loop_agg_is}
\end{figure}

We begin with the aggregate-temporary-impact environment. In that setting, the ex-ante schedule learners converge stably to the corresponding Nash equilibrium. Figure~\ref{fig:open_loop_agg_is} reports the training progression in the \((\mathrm{IS}_0/N,\mathrm{IS}_1/N)\) plane, where the benchmark point is the aggregate-temporary-impact Nash equilibrium. Across runs, the trajectories contract toward the Nash point and remain tightly concentrated in its neighborhood. The main conclusion from this experiment is that, under aggregate temporary impact, removing intra-episode feedback leads learning to recover the competitive schedule stably. A model-based ex-ante schedule-learning variant gives the same qualitative conclusion and is reported in Appendix~\ref{app:deterministic_surrogate_schedule_learning}.

\begin{figure}[!htbp]
    \centering
    \includegraphics[width=0.95\textwidth]{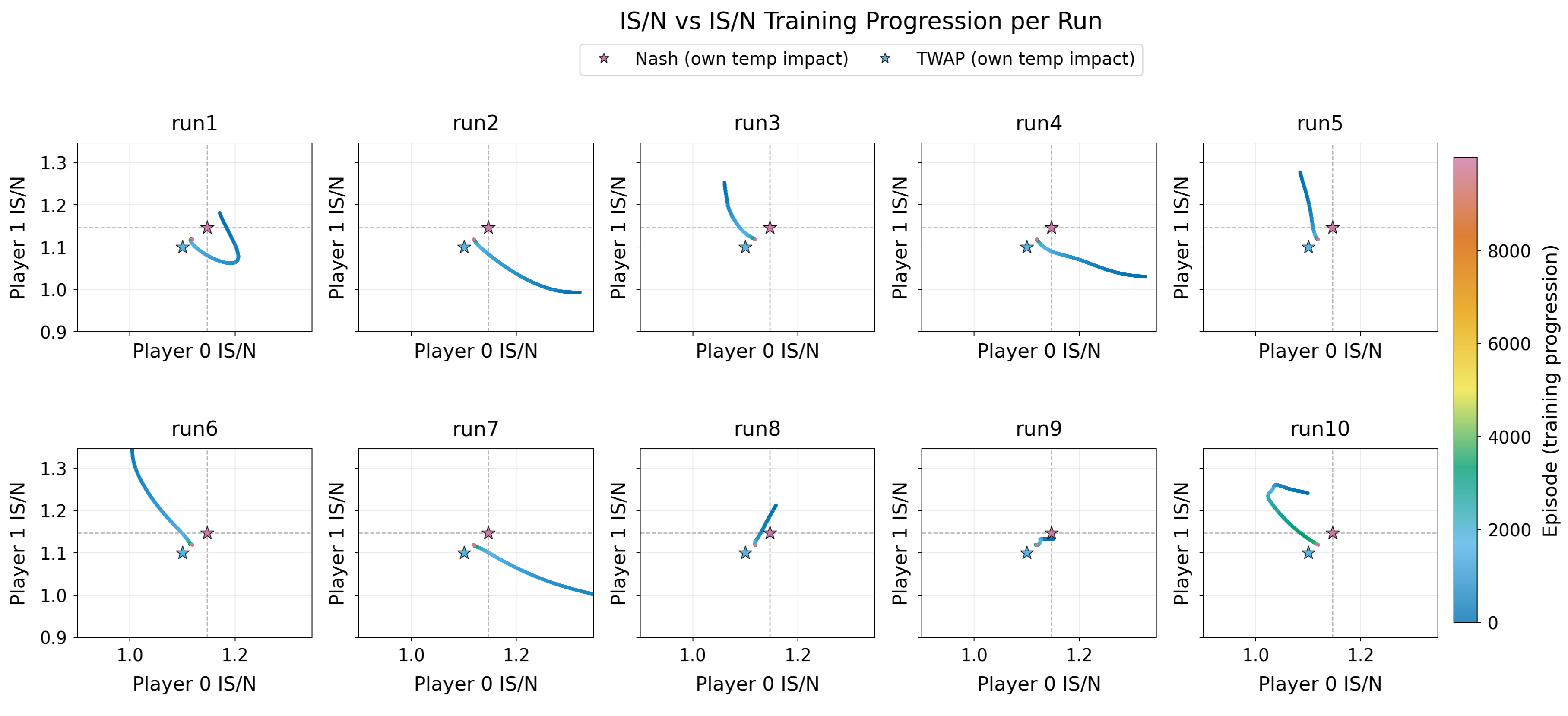}
    \caption{Schedule-learning results in the own-temporary-impact environment.}
    \label{fig:open_loop_own_is}
\end{figure}

The own-temporary-impact environment produces a more subtle picture. If one evaluates the learned outcomes relative to the relevant Nash benchmark, then the agents appear to converge to a supra-competitive region, in the same broad sense as in \citet{lillo2024deviations}: the achieved implementation shortfalls lie below the Nash point. Figure~\ref{fig:open_loop_own_is} illustrates this for the model-free agent. At first sight, this might suggest that schedule learning in the own-temporary-impact environment generates genuinely supra-competitive liquidation trajectories, even without intra-episode feedback.

\begin{figure}[!htbp]
    \centering
    \includegraphics[width=0.95\textwidth]{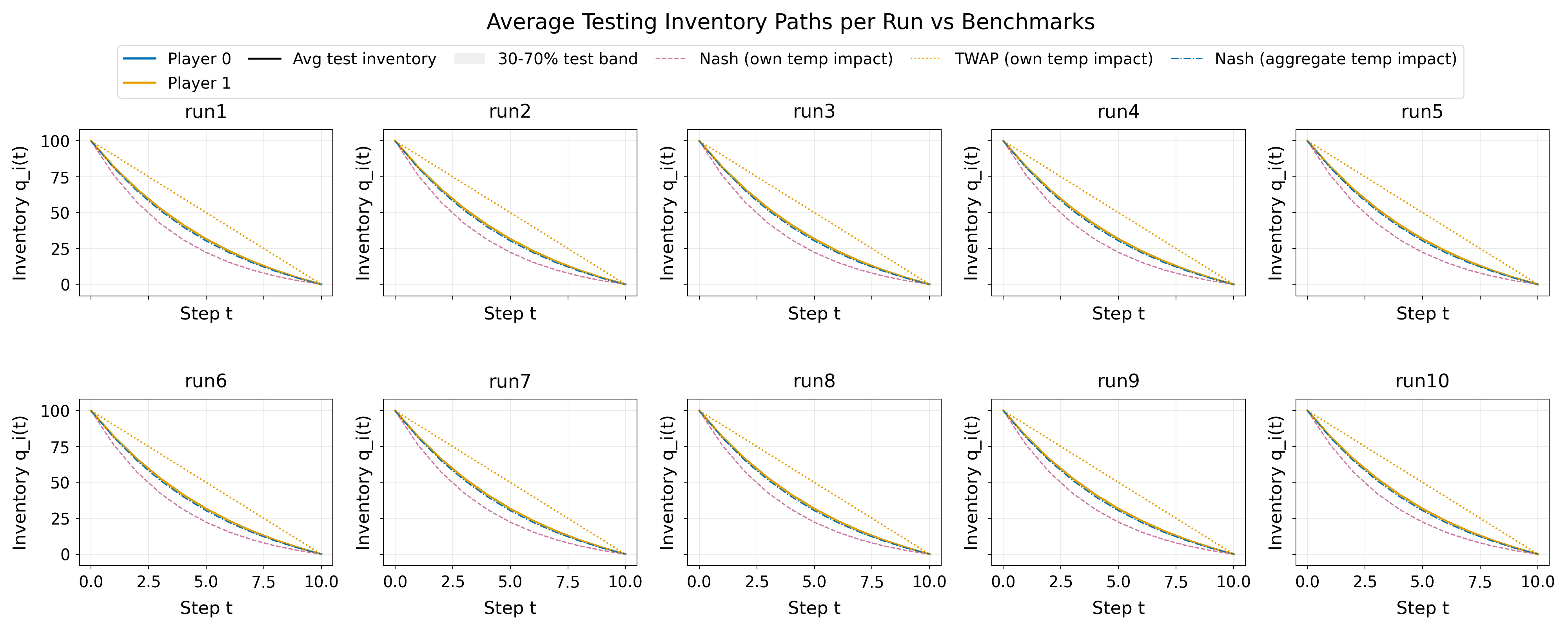}
    \caption{Average testing inventory paths in the own-temporary-impact environment.}
    \label{fig:open_loop_own_paths}
\end{figure}

However, this interpretation changes once the learned inventory paths are compared directly with the benchmark schedules. Figure~\ref{fig:open_loop_own_paths} shows that the learned paths are not drifting toward TWAP, nor toward an arbitrary cooperative schedule. Instead, they lie very close to the aggregate-temporary-impact Nash path, even though the agents are trained in the own-temporary-impact environment. Thus, what appears in the implementation shortfall plane as a supra-competitive outcome relative to the Nash Equilibrium benchmark is, at the path level, much more structured: the learners behave as if the aggregate-temporary-impact Nash schedule remains the relevant competitive reference.

We interpret this as a limited-information phenomenon. In the schedule-learning setting, agents choose complete liquidation trajectories at the beginning of the episode and do not condition on the realized intra-episode price path. Moreover, the aggregate- and own-temporary-impact environments share the same state dynamics,
\[
q_{t+1}^{(i)} = q_t^{(i)} - v_t^{(i)},
\qquad
S_{t+1}
=
S_t - \kappa\left(v_t^{(1)} + v_t^{(2)}\right)
+ \sigma\sqrt{\tau}\,\xi_t,
\]
whereas the difference between the two formulations appears only in the execution price and hence in the instantaneous payoff:
\[
\widetilde S_t^{\mathrm{agg}}
=
S_{t-1}
-
a\frac{v_t^{(1)}+v_t^{(2)}}{\tau},
\qquad
\widetilde S_t^{(i),\mathrm{own}}
=
S_{t-1}
-
a\frac{v_t^{(i)}}{\tau}.
\]
Equivalently, the own-temporary-impact convention affects realized execution revenues, but it does not alter the transition law of the state variables observed along the rollout. The dominant dynamic signal therefore remains the common permanent-impact channel. In this sense, the agents are not fully able to take informed decisions regarding the benchmark distinction through the state evolution itself: the own-temporary-impact detail enters only contemporaneously through payoffs, and does not generate a distinct state-transition signal that would naturally guide learning toward a different path.

This interpretation helps explain why the own-temporary-impact experiment can look supra-competitive without the agents appearing to discover a genuinely new cooperative schedule. A researcher who evaluates outcomes against the Nash benchmark will classify the result as supra-competitive. Yet the path comparison indicates that the learners are instead organizing around the aggregate-temporary-impact Nash schedule. Put differently, the agents do not appear to internalize the own-temporary-impact benchmark as the relevant competitive reference. Rather, under the informational restrictions created by removing intra-episode feedback, they behave as if the aggregate-temporary-impact Nash schedule is the effective benchmark.

\begin{figure}[!htbp]
    \centering
    \includegraphics[width=0.95\textwidth]{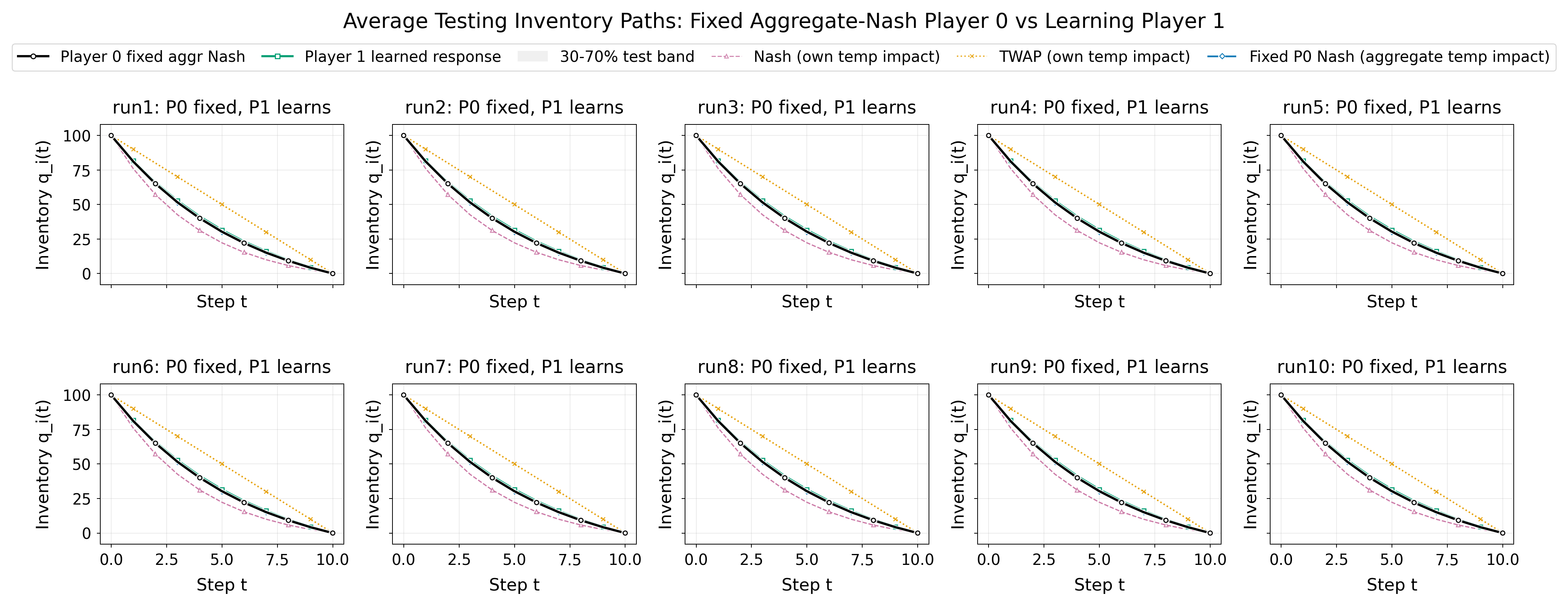}
    \caption{Own-temp experiment with one player fixed at the aggregate-temporary-impact Nash schedule.}
    \label{fig:open_loop_fixed_agg_response}
\end{figure}

To probe this interpretation further, we ran an additional experiment in the own-temporary-impact environment. One player was fixed exogenously at the aggregate-temporary-impact Nash schedule, while the other player was allowed to learn. The purpose was to test how the learned aggregate-Nash-like path is treated by the learning dynamics. If the learning agent interpreted that path as a supra-competitive opportunity, then one would expect it to deviate from the fixed player's schedule in order to exploit the situation. If, instead, it treated that path as the relevant competitive benchmark, then it should remain close to the same schedule. Figure~\ref{fig:open_loop_fixed_agg_response} shows that the latter is what happens: the learning player converges to a response identical to the fixed aggregate-temporary-impact Nash path. We obtain the same qualitative conclusion for the model-based agent.

This auxiliary experiment reinforces the interpretation above. From the perspective of the schedule learner, the aggregate-temporary-impact Nash schedule appears to function as an effective equilibrium reference, even inside the own-temporary-impact environment. The learner does not respond to a fixed aggregate-Nash opponent as if facing an obviously exploitable supra-competitive profile; instead, it remains close to that same path
\section{Baseline State-Dependent Multi-Agent Deep Reinforcement Learning Experiment}
\label{sec:baseline-experiment}

We now allow agents to receive intra-episode feedback from the execution environment. Unlike the schedule learners in Section~\ref{sec:open_loop_rl}, the agents studied here observe the evolving state during the liquidation episode and can condition their subsequent actions on information generated by the interaction. This allows us to evaluate whether state-dependent feedback alone is sufficient to move learned behavior away from the competitive benchmark. The experiment follows the DDQN framework of \citet{lillo2024deviations}, but with several design choices adjusted to match the environment studied in this paper.

First, agents submit their trades simultaneously at each timestep. This removes any possible within-slice information leakage. Second, exploration is distributed equivalently over the admissible action set rather than around the TWAP schedule, so that TWAP is not built directly into the exploration process. Third, given the benchmark analysis and the schedule-learning results above, we use the aggregate-temporary-impact environment as the baseline setting.

 Closely to \citet{lillo2024deviations}, we consider the symmetric two-agent case and train two independent DDQN agents jointly in the Almgren--Chriss environment for \(10{,}000\) episodes. Training is repeated over \(10\) independent random seeds; after each run, the learned pair is evaluated jointly over \(500\) out-of-sample test episodes using greedy action selection.

The per-step reward is implemented as
\[
r_t^{(i)}
=
-\frac{S_0 q_0^{(i)}}{N}
+\widetilde S_t\,v_t^{(i)}
-a\bigl(v_t^{(i)}\bigr)^2.
\]
The first term allocates the initial mark-to-market value \(S_0 q_0^{(i)}\) uniformly across the \(N\) trading periods, while the remaining terms represent the revenue obtained from the executed trade and the additional own temporary-impact penalty used in the learning objective. Consequently, maximizing cumulative reward is equivalent, up to the sign and normalization convention, to minimizing implementation shortfall in the discrete execution environment.

Each agent \(i\) learns its own state--action value function \(Q_{\theta_i}(s,a)\) from a replay buffer, using as input its current price, own remaining inventory, time index, and candidate action, after normalization. The value function is approximated by a fully connected neural network with five hidden layers of width \(128\), SiLU activations, and a scalar output. Exploration follows an \(\varepsilon\)-greedy rule over admissible discrete trades,
\[
a_t^{(i)}=
\begin{cases}
\text{random admissible action}, & \text{with probability } \varepsilon_t,\\[2mm]
\arg\max_{a\in\mathcal A_t^{(i)}} Q_{\theta_i}(s_t^{(i)},a), & \text{with probability } 1-\varepsilon_t,
\end{cases}
\]
where \(\varepsilon_t\) decays multiplicatively from \(1\) to \(0.05\) and is then held fixed at its floor. The DDQN target for agent \(i\) is
\[
y_t^{(i)}
=
r_t^{(i)}
+
(1-d_t)\gamma\,
Q_{\bar\theta_i}\!\left(s_{t+1}^{(i)},
\arg\max_{a'\in\mathcal A_{t+1}^{(i)}} Q_{\theta_i}(s_{t+1}^{(i)},a')\right),
\]
and parameters are estimated by minimizing the mean-squared Bellman error
\[
\mathcal L(\theta_i)
=
\frac{1}{|\mathcal B|}
\sum_{(s,a,r,s',d)\in\mathcal B}
\left(
Q_{\theta_i}(s,a)-y
\right)^2,
\]
with the target network updated by Polyak averaging,
\[
\bar\theta_i \leftarrow \tau \theta_i + (1-\tau)\bar\theta_i.
\]
Table~\ref{tab:ddqn_replication_hyperparams} reports the hyperparameters used in this baseline replication. It should be noted that we extended our experiments with more training episodes and did not observe any significant difference.

\begin{table}[!htbp]
\centering
\caption{Hyperparameters for the baseline two-agent DDQN replication.}
\label{tab:ddqn_replication_hyperparams}
\begin{tabular}{ll|ll}
\toprule
\textbf{Hyperparameter} & \textbf{Value} & \textbf{Hyperparameter} & \textbf{Value} \\
\midrule
Number of agents \(K\) & \(2\) & Discount factor \(\gamma\) & \(1.0\) \\
Initial inventories \(q_0\) & \((100,100)\) & Optimizer & Adam \\
Initial price \(S_0\) & \(10\) & Learning rate & \(2\times 10^{-4}\) \\
Horizon \(T\) & \(10\) & Replay-memory capacity & \(15{,}000\) \\
Time step \(\Delta t\) & \(1\) & Mini-batch size & \(128\) \\
Number of trading periods \(N\) & \(10\) & Target-network update rate \(\tau\) & \(5\times 10^{-3}\) \\
Permanent impact \(\kappa\) & \(0.001\) & Exploration rate \(\varepsilon_{\mathrm{start}}\) & \(1.0\) \\
Temporary impact \(a\) & \(0.002\) & Minimum exploration rate \(\varepsilon_{\mathrm{min}}\) & \(0.05\) \\
Volatility \(\sigma\) & \(10^{-9}\) & Fraction of training until \(\varepsilon_{\min}\) & \(40\%\) \\
Training episodes & \(10{,}000\) & Network depth & \(5\) hidden layers \\
Test episodes & \(500\) & Hidden width & \(128\) units per layer \\
Independent runs & \(10\) & Activation & SiLU \\
\bottomrule
\end{tabular}
\end{table}

\begin{figure}[!htbp]
    \centering

    \begin{subfigure}{0.7\textwidth}
        \centering
        \includegraphics[width=\textwidth]{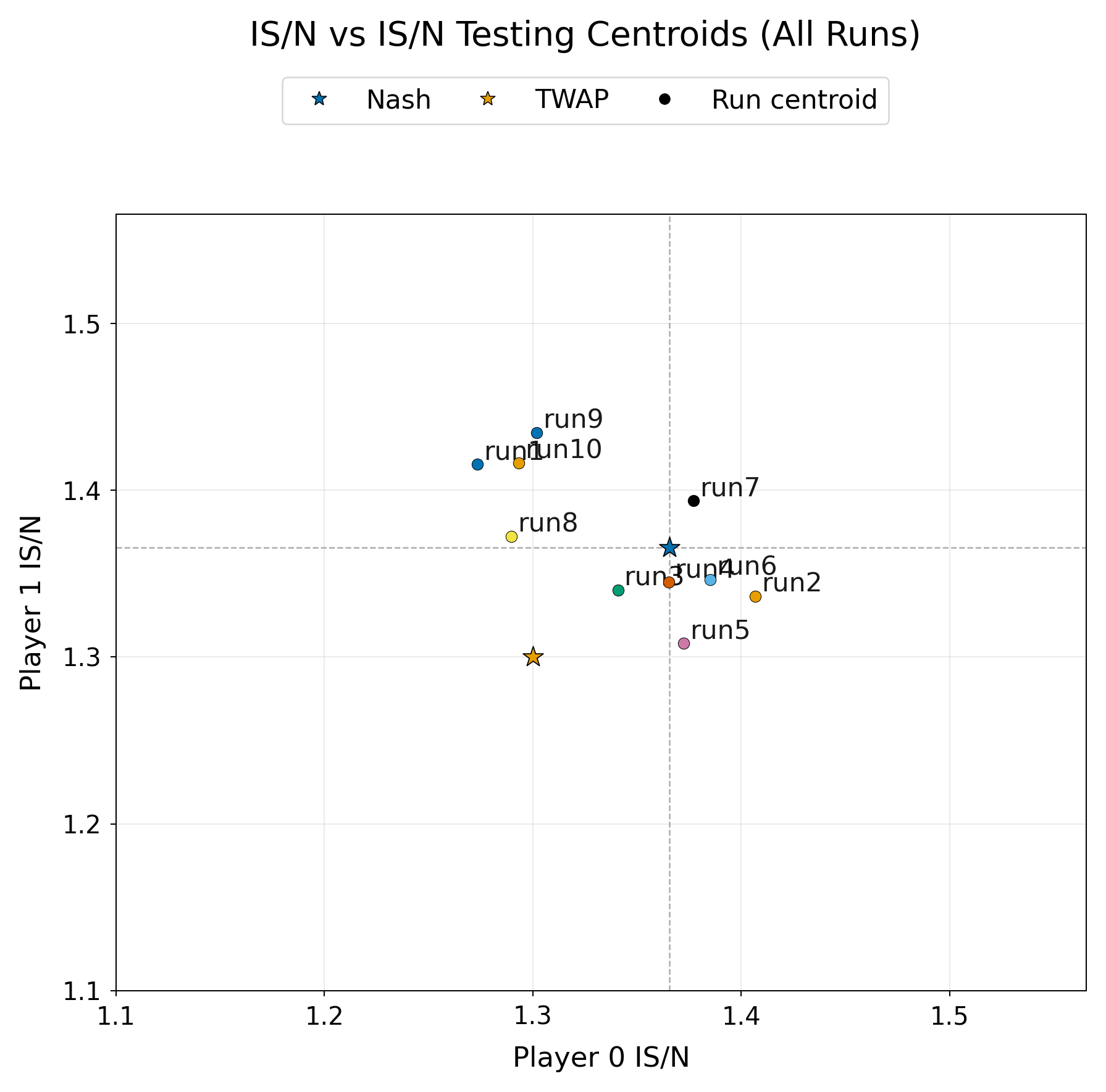}
        \caption{}
        \label{fig:nash_training_group_a}
    \end{subfigure}

    \vspace{0.8em}

    \begin{subfigure}{1\textwidth}
        \centering
        \includegraphics[width=\textwidth]{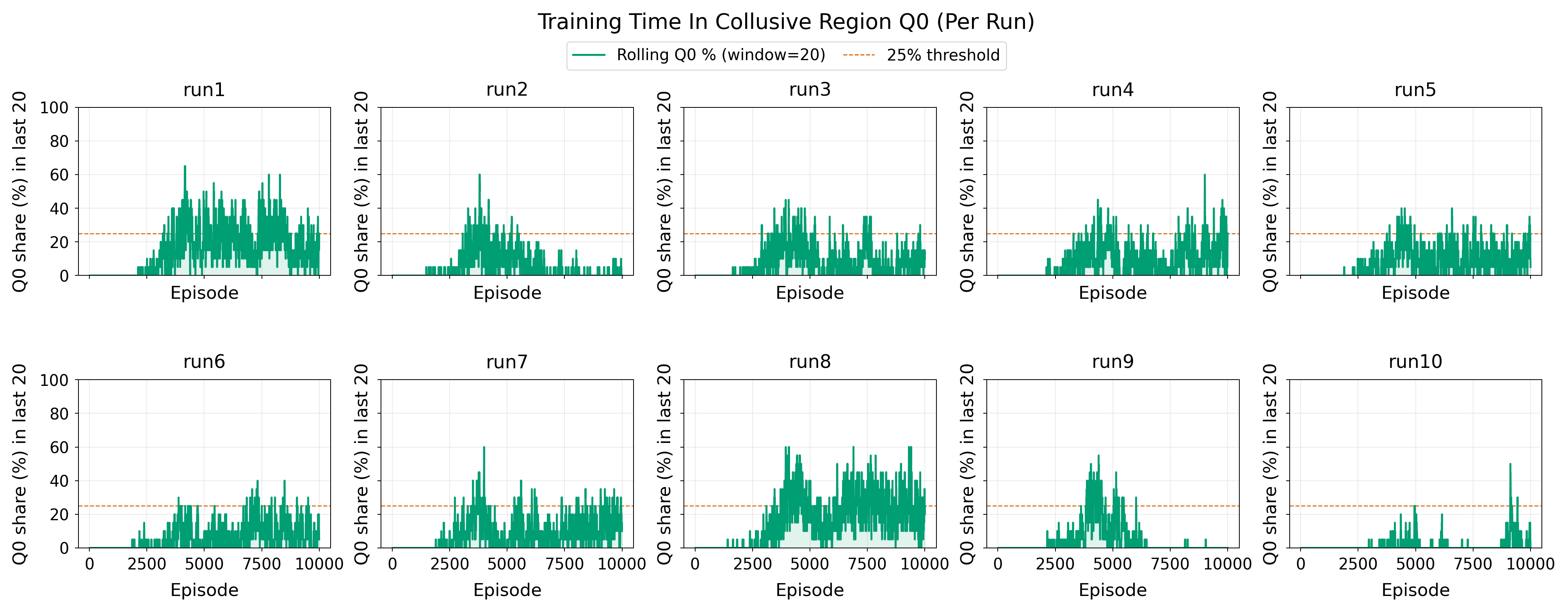}
        \caption{}
        \label{fig:nash_training_group_b}
    \end{subfigure}

    \vspace{0.8em}

    \caption{Baseline DDQN experiment with intra-episode feedback. \textbf{(a)} Final testing centroids relative to the aggregate-temp Nash and TWAP benchmarks. \textbf{(b)} Rolling \(20\)-episode share of training episodes in the supra-competitive region relative to the aggregate-temp Nash benchmark.}
    \label{fig:nash_training_group}
\end{figure}

Figure~\ref{fig:nash_training_group} summarizes the resulting behavior at both test time and during training. Panel~\textbf{(a)} reports the final testing centroids in the \((\mathrm{IS}_0/N,\mathrm{IS}_1/N)\) plane against the aggregate-temporary-impact Nash benchmark and the TWAP benchmark. The Nash point partitions the plane into four regions, with the south-west region corresponding to outcomes in which both players achieve lower implementation shortfall than the competitive Nash benchmark. We therefore refer to this region as the supra-competitive region. The testing centroids remain clustered around the Nash benchmark rather than moving systematically to the south-west of it. Some runs lie marginally below the Nash point for one player, but there is no stable pattern in which both agents jointly achieve lower implementation shortfall than Nash. The outcomes therefore remain much closer to the competitive benchmark than to the TWAP benchmark.

Panel~\textbf{(b)} reports, for each run and over a rolling window of \(20\) episodes, the fraction of training episodes that end in the supra-competitive region. Since the Nash point partitions the plane into four quadrants, the \(25\%\) line serves as a natural reference level: it is the share one would expect if the implementation-shortfall outcomes were fluctuating randomly around the Nash equilibrium. The rolling share fluctuates during training, but it does not converge to persistently high values. In most runs, it remains below or around the \(25\%\) reference level. Thus, allowing intra-episode feedback through the baseline DDQN state representation is not, by itself, sufficient to produce stable supra-competitive behavior in this simultaneous aggregate-impact environment.

A natural question is whether the baseline DDQN architecture is in fact able to exploit the price signal available in the state. To diagnose whether the learned Q-function actually depends on price, we use integrated gradients as a post hoc attribution method. For a probe state \(s\), let
\[
a^*(s)\in \arg\max_{a\in\mathcal A(s)} Q_{\theta}(s,a)
\]
denote the greedy action under the current network, and define the normalized feature vector
\[
x(s)
=
\bigl(S^{\mathrm{norm}}(s),\, I^{\mathrm{norm}}(s),\, t^{\mathrm{norm}}(s),\, a^*(s)^{\mathrm{norm}}\bigr)\in\mathbb R^4.
\]
We explain the scalar quantity
\[
f_{\theta}(s)\coloneqq Q_{\theta}\!\bigl(s,a^*(s)\bigr)
\]
relative to the baseline input
\[
x_0(s)=\bigl(0,0,0,a^*(s)^{\mathrm{norm}}\bigr),
\]
which corresponds to initial price, initial inventory, and initial time, while keeping the explained action fixed. For feature \(j\), the integrated-gradient attribution is
\begin{equation}
\mathrm{IG}_j(x;x_0)
=
(x_j-x_{0,j})
\int_0^1
\frac{\partial f_{\theta}\!\bigl(x_0+\alpha(x-x_0)\bigr)}{\partial x_j}
\,d\alpha .
\label{eq:ig_def}
\end{equation}
In practice, we approximate \eqref{eq:ig_def} by a Riemann sum with \(m\) interpolation steps:
\begin{equation}
\widehat{\mathrm{IG}}_j(x;x_0)
=
(x_j-x_{0,j})
\frac{1}{m}
\sum_{\ell=1}^{m}
\frac{\partial f_{\theta}\!\bigl(x_0+\frac{\ell}{m}(x-x_0)\bigr)}{\partial x_j}.
\label{eq:ig_approx}
\end{equation}

\begin{figure}[!htbp]
    \centering
    \begin{subfigure}{0.7\textwidth}
        \centering
        \includegraphics[width=\textwidth]{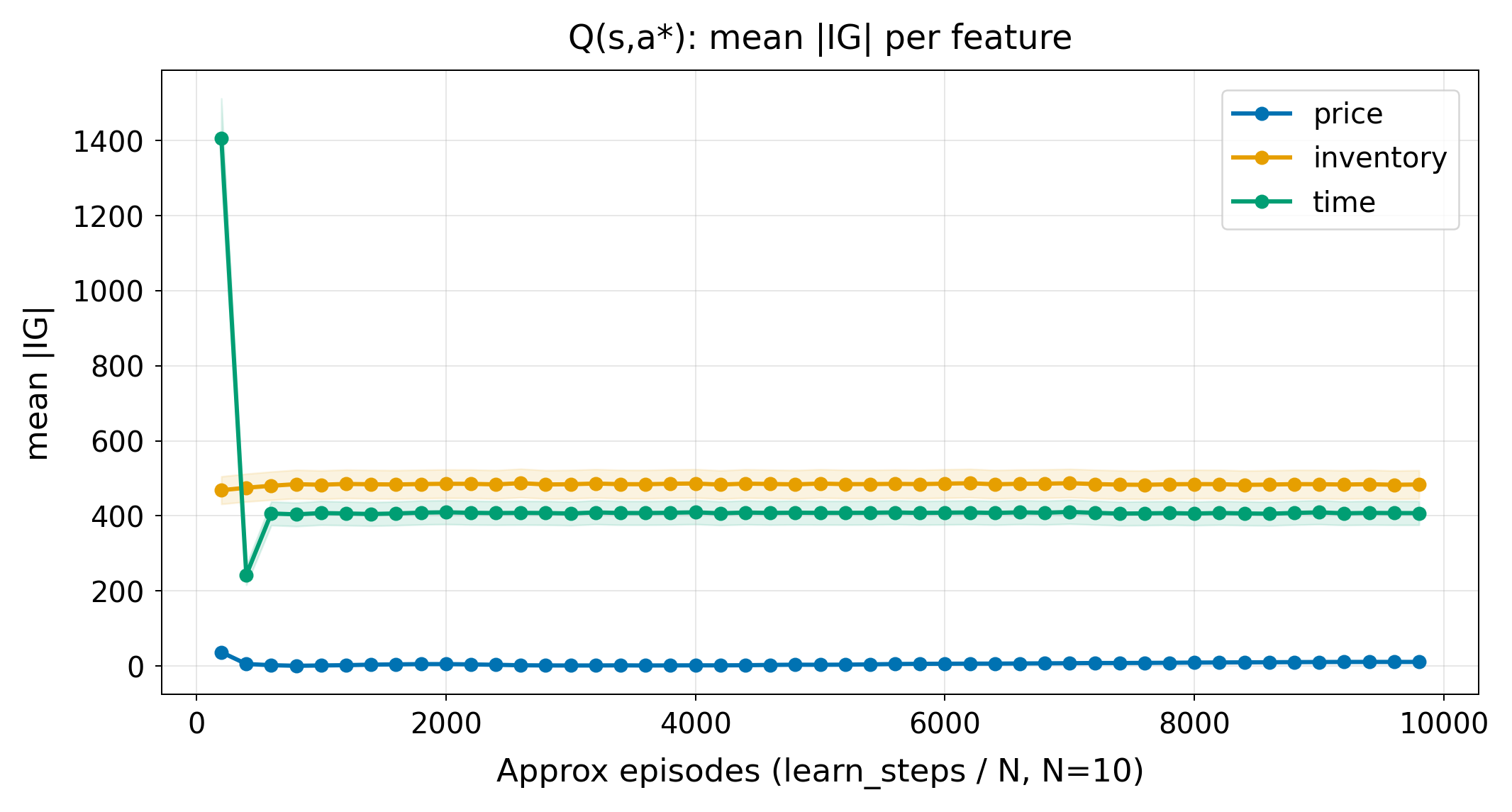}
    \end{subfigure}
    \caption{Integrated-gradient diagnostics for the baseline DDQN architecture in an illustrative run. The figure reports the mean absolute attribution \(A_j\) for price, inventory, and time over a fixed grid of probe states. After the initial phase of learning, the price attribution remains close to zero, while inventory and time account for essentially all of the variation in \(Q(s,a^*)\).}
    \label{fig:ig_run4_baseline}
\end{figure}

We evaluate these attributions on a fixed off-policy probe set \(\mathcal G\), constructed from a grid over prices, inventories, and times, and summarize feature importance by the mean absolute attribution
\begin{equation}
A_j
=
\frac{1}{|\mathcal G|}
\sum_{s\in\mathcal G}
\left|
\widehat{\mathrm{IG}}_j\bigl(x(s);x_0(s)\bigr)
\right|.
\label{eq:ig_summary}
\end{equation}
Figure~\ref{fig:ig_run4_baseline} reports \(A_j\) over training for an illustrative run. The pattern is striking: after a short initial transient, the attribution assigned to price collapses to a level close to zero, whereas inventory and time dominate throughout learning. Thus, under the baseline architecture, the learned Q-function behaves essentially as a liquidation schedule indexed by remaining inventory and time, with negligible dependence on the contemporaneous price. This does \emph{not} imply that price is economically irrelevant; rather, it shows that the baseline network does not learn to use that signal in a meaningful way.

The fact that the baseline network assigns little importance to price does not mean that the agents behave as if they were solving an isolated single-agent execution problem. If they ignored the strategic nature of the environment altogether, then in the symmetric risk-neutral setting one would expect them to move toward the TWAP schedule, which is optimal for a single trader and coincides with the Pareto-efficient benchmark in the two-player game. Instead, the learned policies remain much closer to the Nash benchmark. This suggests that the agents do learn that they are interacting with another trader, most likely through the reward signal induced by aggregate market impact. However, because the learned value function does not use the contemporaneous price in a meaningful way, the agents cannot exploit price-based information as a coordination channel for reaching persistent supra-competitive outcomes. This motivates the next experiment, where we modify the Q-network architecture to make the price coordinate more salient and test whether stronger price conditioning is sufficient to move learned behavior away from Nash.
\section{The Price-Conditioned Agents}
\label{subsec:price_conditioned_network}

To test whether the weak price dependence of the baseline DDQN is mainly architectural, we repeat the benchmark experiment of the previous section with a modified Q-network designed to strengthen the price channel. We keep the same training protocol, hyperparameters, and evaluation procedure as before, and change only the network architecture. The modified network combines three mechanisms: a learnable rescaling of the price coordinate, feature-wise linear modulation (FiLM) of each residual block following \citet{perez2018film}, and an additive price-specific skip head.

Let the normalized input be
\[
x=(p,i,\tau,a)\in\mathbb R^4,
\]
where \(p=S^{\mathrm{norm}}\), \(i=I^{\mathrm{norm}}\), \(\tau=t^{\mathrm{norm}}\), and \(a=a^{\mathrm{norm}}\). First, the price coordinate is rescaled by a learned scalar \(c_p>0\),
\[
\widetilde p = c_p\, p,
\qquad
\widetilde x = (\widetilde p,i,\tau,a).
\]
The base trunk then produces an initial hidden representation
\begin{equation}
h^{(0)} = \phi_{\mathrm{base}}(\widetilde x),
\label{eq:base_trunk}
\end{equation}
where \(\phi_{\mathrm{base}}\) is a two-layer MLP with SiLU activations. In parallel, a price-only embedding is computed as
\begin{equation}
c = \phi_{\mathrm{price}}(\widetilde p),
\label{eq:price_embedding}
\end{equation}
with \(\phi_{\mathrm{price}}\) another two-layer SiLU MLP.

For each residual block \(\ell=1,\dots,L\), the price embedding generates FiLM parameters
\begin{equation}
(\gamma^{(\ell)},\beta^{(\ell)})
=
W^{(\ell)}_{\mathrm{film}} c + b^{(\ell)}_{\mathrm{film}},
\label{eq:film_params}
\end{equation}
where \(\gamma^{(\ell)},\beta^{(\ell)}\in\mathbb R^H\). To stabilize training, the multiplicative modulation is bounded as
\begin{equation}
\widetilde\gamma^{(\ell)} = \lambda_{\mathrm{film}} \tanh\!\bigl(\gamma^{(\ell)}\bigr),
\label{eq:film_gamma}
\end{equation}
and the hidden state is modulated feature-wise according to
\begin{equation}
\widetilde h^{(\ell-1)}
=
h^{(\ell-1)}\odot \bigl(1+\widetilde\gamma^{(\ell)}\bigr)
+
\beta^{(\ell)}.
\label{eq:film_modulation}
\end{equation}
The residual update is then
\begin{equation}
h^{(\ell)}
=
h^{(\ell-1)}
+
\rho\,
B^{(\ell)}\!\bigl(\widetilde h^{(\ell-1)}\bigr),
\label{eq:residual_update}
\end{equation}
where \(B^{(\ell)}\) is a LayerNorm--Linear--SiLU--Linear block and \(\rho>0\) is a residual scaling constant.

Finally, the Q-value is the sum of a trunk head and a price skip head:
\begin{equation}
Q_{\theta}(x)
=
w_h^\top h^{(L)} + b_h
+
\eta\,
\bigl(w_p^\top c + b_p\bigr),
\label{eq:price_conditioned_q}
\end{equation}
where \(\eta>0\) is a learned output scale. This architecture does not guarantee that the agent will use price in an economically meaningful way, but it creates an explicit and much stronger pathway through which price can influence the learned value function.

\begin{figure}[!htbp]
    \centering
    \begin{subfigure}{0.7\textwidth}
        \centering
        \includegraphics[width=\textwidth]{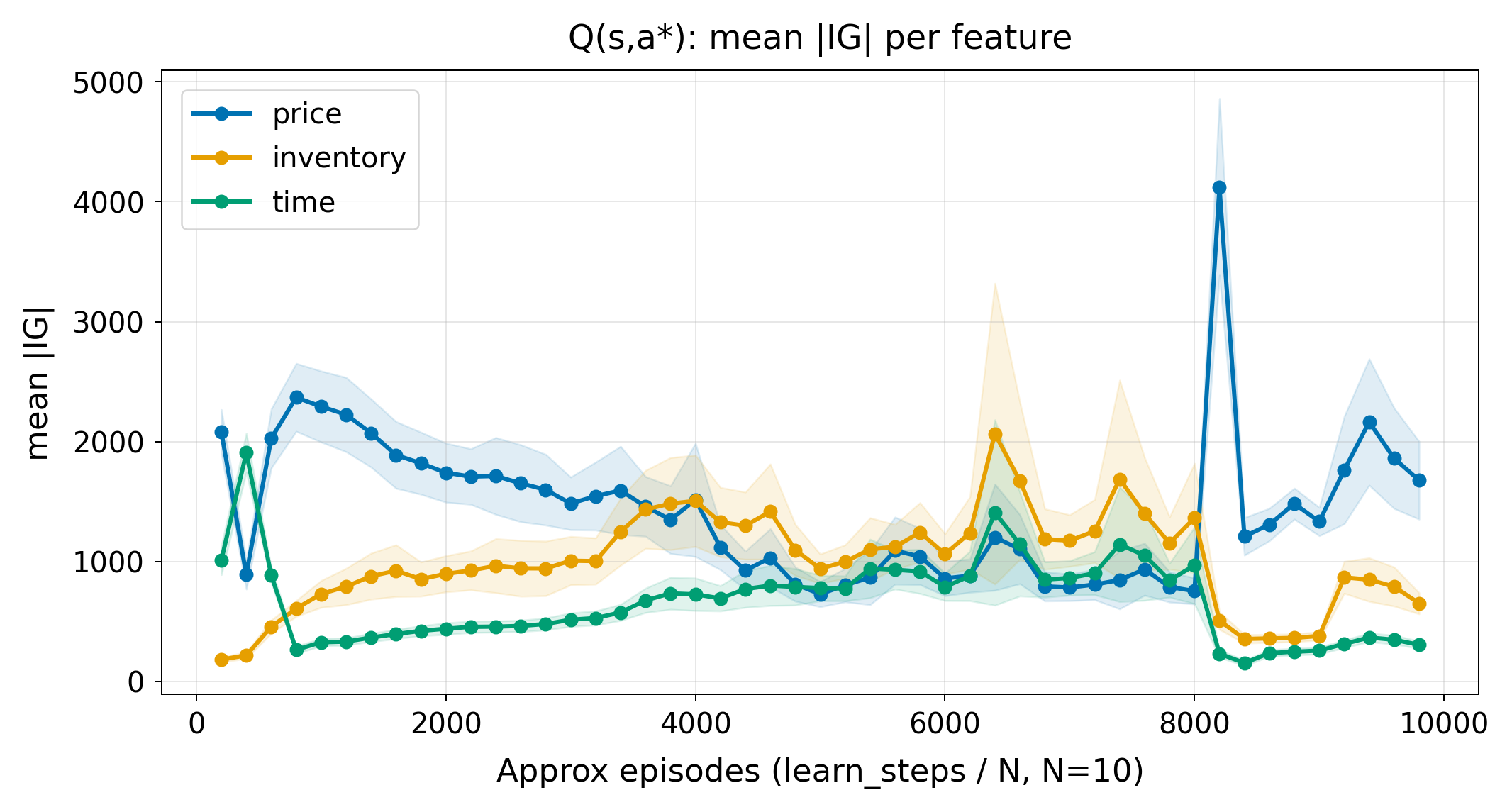}
    \end{subfigure}
    \caption{Integrated-gradient diagnostics for the price conditioned DDQN architecture in an illustrative run. The price attribution is now significant, contributing among inventory and time to the variation in \(Q(s,a^*)\).}
    \label{fig:price_conditioned_ig}
\end{figure}

We then repeat the same experiment as in the previous section, with one adjustment to the training horizon. The two agents are trained jointly in the same environment, tested jointly over \(500\) out-of-sample episodes, and the exercise is replicated over \(10\) independent runs. Because the price-conditioned architecture appears to require longer training to stabilize, we train it for \(20{,}000\) episodes rather than \(10{,}000\). To keep the amount of exploration comparable to the baseline experiment, we keep fixed the absolute number of exploration-dominated episodes: \(\epsilon\) still reaches its minimum value \(\epsilon_{\min}\) after \(4{,}000\) episodes, which corresponds to reducing the fraction of training required to reach \(\epsilon_{\min}\) from \(40\%\) to \(20\%\). The resulting integrated-gradient diagnostics are shown in Figure~\ref{fig:price_conditioned_ig}.

\begin{figure}[!htbp]
    \centering

    \begin{subfigure}{0.7\textwidth}
        \centering
        \includegraphics[width=\textwidth]{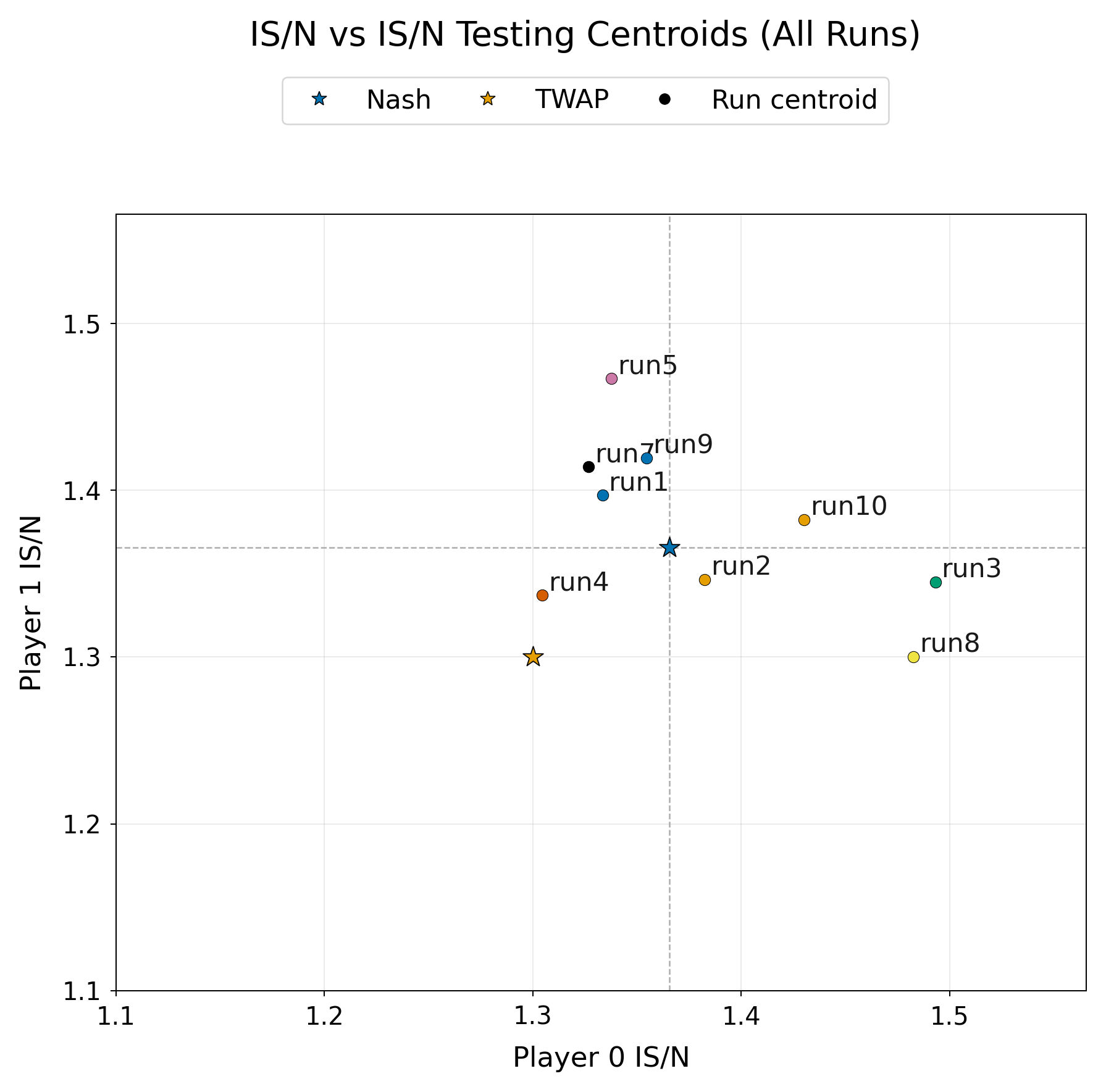}
        \caption{}
        \label{fig:nash_training_group_a_price}
    \end{subfigure}

    \vspace{0.8em}

    \begin{subfigure}{1\textwidth}
        \centering
        \includegraphics[width=\textwidth]{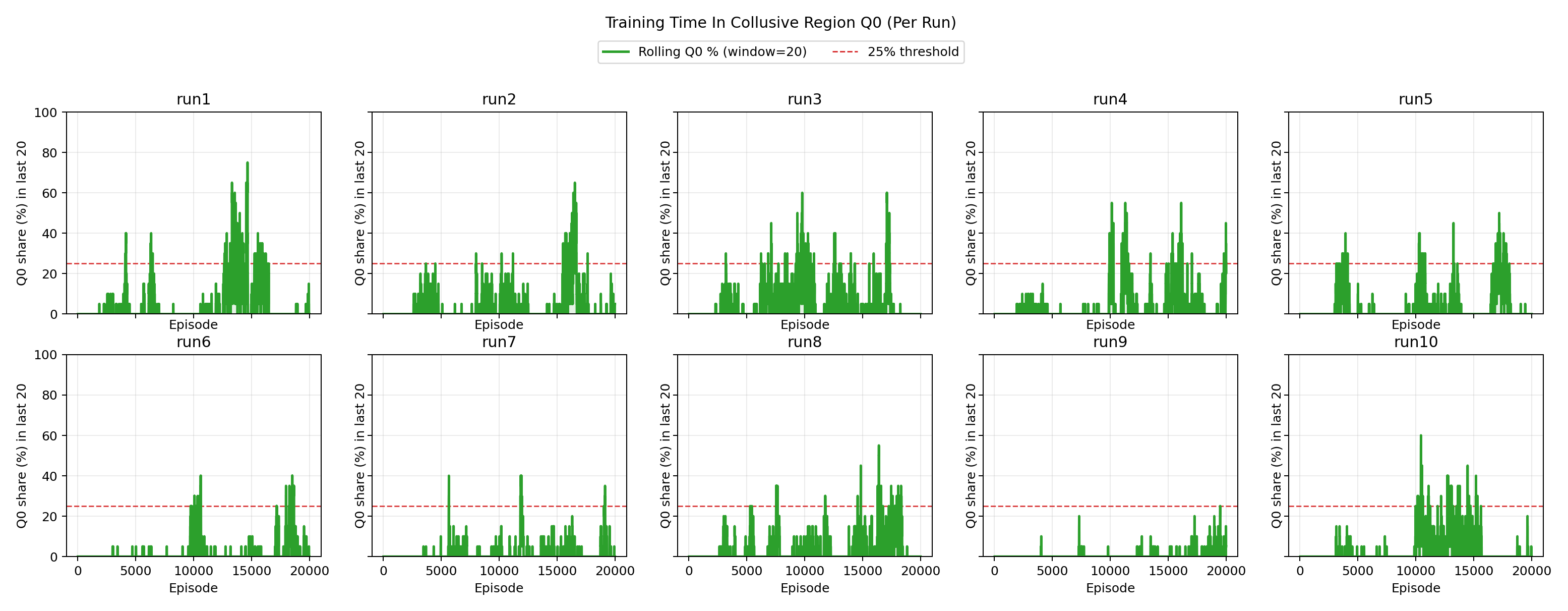}
        \caption{}
        \label{fig:nash_training_group_b_price}
    \end{subfigure}    
    \caption{Price-conditioned DDQN replication. \textbf{(a)} Final testing centroids relative to the continuous-time and grid-implemented Nash and TWAP benchmarks. \textbf{(b)} Rolling \(20\)-episode share of training episodes in the collusive region under the discrete benchmark.}
    \label{fig:price_conditioned_centroids}
\end{figure}

However, the price-conditioned architecture still does not produce a systematic shift toward supra-competitive outcomes. Figure~\ref{fig:price_conditioned_centroids} summarizes the resulting behavior at both test time and during training. Panel~\textbf{(a)} reports the final testing centroids in the \((\mathrm{IS}_0/N,\mathrm{IS}_1/N)\) plane against the continuous-time and grid-implemented Nash and TWAP benchmarks. Relative to the discrete Nash benchmark, only a limited number of runs fall inside the collusive region, defined here as the south-west quadrant in which both players achieve lower implementation shortfall than the discrete Nash point. Most testing centroids instead remain close to the discrete Nash benchmark, while some lie above it. Thus, even when the architecture is modified so that current price information enters the value function more directly, the resulting learned behavior remains predominantly near-Nash rather than robustly supra-competitive.
In panel~\textbf{(b)}, for each run we compute the rolling share of training episodes that end in the collusive region, using a \(20\)-episode moving window and the discrete Nash benchmark as the reference point. As before, the \(25\%\) line provides a natural baseline because the discrete Nash point partitions the plane into four quadrants. Across runs, the rolling share generally remains below, or only briefly exceeds, that threshold, with occasional spikes that again point to transient instability rather than persistent collusive convergence.

This finding is important for the interpretation of the earlier results. The absence of stable supra-competitive outcomes cannot be attributed solely to the baseline MLP underweighting the price coordinate. Even after strengthening the price channel architecturally, the qualitative pattern remains heterogeneous across runs. This points to a deeper informational limitation: the agents still observe only the current price, not its intra-episode history. The same price level can be reached through different sequences of aggregate order flow, and these sequences may contain information about the liquidation behavior of the other trader that is not recoverable from the current price alone. If coordination or punishment is encoded in \emph{how} the price was reached, rather than only in the static price level itself, then an architecture without explicit memory may still be fundamentally unable to recover the relevant signal. This motivates the next step of the analysis, namely to augment the state representation with price history that can track within-episode dynamics.

\section{The History-Aware Agents}
\label{sec:price-history-memory}

The previous architectures evaluate the candidate action using only contemporaneous state variables. To allow the agent to condition its decision on \emph{how} the current price was reached within the episode, we augment the Q-network with an explicit sequence-processing branch. At each decision time \(t\), for a candidate action \(a\), the network now receives the current core state together with the within-episode history of observed prices and the agent's own past executed actions. Denoting the episode length by \(N\), the state--action input can be written as
\[
z_t(a)
=
\Bigl[
S_t,\; q_t,\; t,\;
p_0,\dots,p_{N-1},\;
u_{-1},u_0,\dots,u_{N-2},\;
m_0,\dots,m_{N-1},\;
a
\Bigr],
\]
where \(p_\ell\) is the observed price at step \(\ell\), \(u_{\ell-1}\) is the agent's own previously executed action aligned with that step (with \(u_{-1}=0\) by convention), and \(m_\ell\in\{0,1\}\) is a validity mask indicating whether position \(\ell\) has already been observed within the current episode.

The sequence branch forms a token at each episode position from the normalized price and lagged action,
\[
x_\ell
=
\begin{bmatrix}
p_{\ell,\mathrm{norm}} \\
u_{\ell-1,\mathrm{norm}}
\end{bmatrix},
\qquad \ell=0,\dots,N-1,
\]
projects it into a \(d\)-dimensional latent space,
\[
e_\ell = W_{\mathrm{tok}} x_\ell + b_{\mathrm{tok}},
\]
and adds positional encodings:
\[
\widetilde e_\ell = e_\ell + \pi_\ell .
\]
The resulting sequence is passed through a Transformer encoder with masked self-attention,
\[
H = \mathrm{Enc}\!\left(\widetilde e_0,\dots,\widetilde e_{N-1};\,m_0,\dots,m_{N-1}\right),
\]
so that the representation at each position may attend to all previously valid positions in the episode. We then aggregate the encoded sequence by masked average pooling:
\[
h_t^{\mathrm{seq}}
=
\frac{\sum_{\ell=0}^{N-1} m_\ell H_\ell}{\sum_{\ell=0}^{N-1} m_\ell}.
\]

In parallel, the current static state and candidate action are processed through a separate MLP branch:
\[
h_t^{\mathrm{stat}}
=
\phi\!\left(
S_{t,\mathrm{norm}},
q_{t,\mathrm{norm}},
t_{\mathrm{norm}},
a_{\mathrm{norm}}
\right).
\]
The final Q-value is produced by fusing the sequential and static representations,
\[
Q_\theta(z_t(a))
=
\psi\!\left(
\bigl[h_t^{\mathrm{seq}};\,h_t^{\mathrm{stat}}\bigr]
\right),
\]
where \(\psi\) is a feed-forward head. In the implementation used here, the sequence encoder uses \(d=64\), \(2\) attention heads, and \(2\) encoder layers, while the static branch and output head are standard GELU MLPs. Relative to the previous experiments, this is the key architectural change: the value function can now depend not only on the current price level, but also on the intra-episode trajectory of prices and past actions.

\begin{figure}[!htbp]
    \centering

    \begin{subfigure}{0.7\textwidth}
        \centering
        \includegraphics[width=\textwidth]{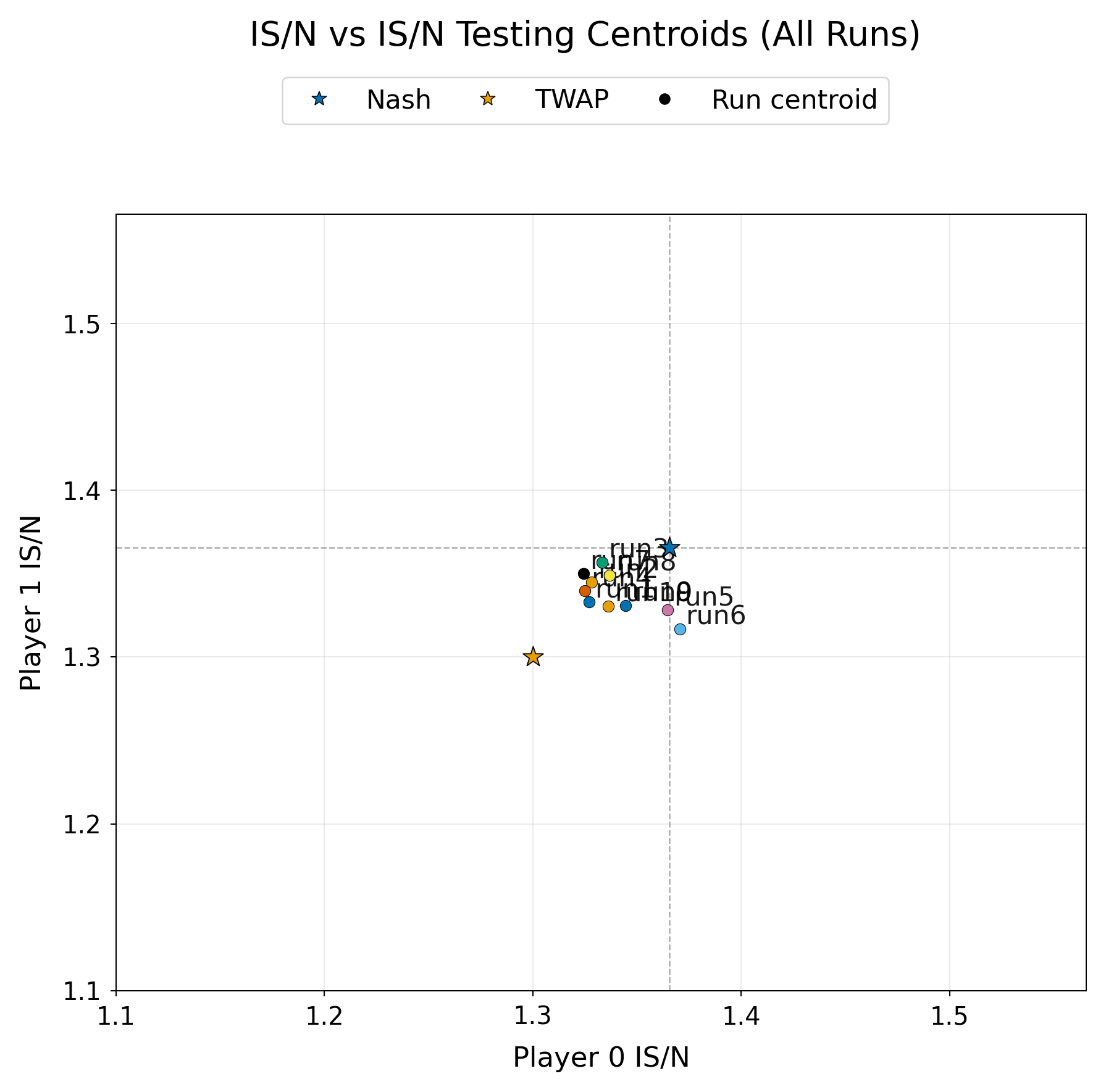}
        \caption{}
        \label{fig:history_attention_group_a}
    \end{subfigure}

    \vspace{0.8em}

    \begin{subfigure}{1\textwidth}
        \centering
        \includegraphics[width=\textwidth]{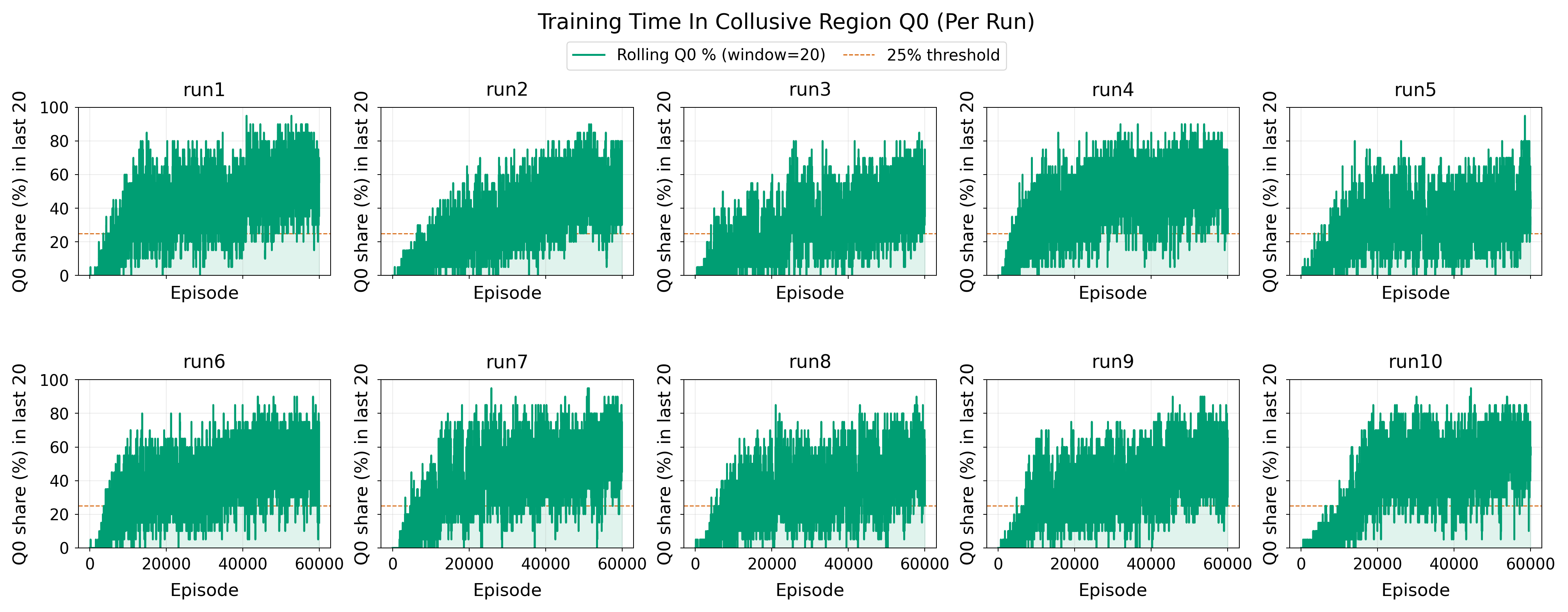}
        \caption{}
        \label{fig:history_attention_group_b}
    \end{subfigure}
    \caption{History-aware DDQN replication. \textbf{(a)} Final testing centroids relative to the Nash and TWAP benchmarks. \textbf{(b)} Rolling \(20\)-episode share of training episodes in the supra-competitive region under the competitive benchmark.}
    \label{fig:history_attention_group}
\end{figure}

Figure~\ref{fig:history_attention_group} shows that once intra-episode history is incorporated into the state representation, the qualitative picture changes substantially. Panel~\textbf{(a)} reports the final testing centroids in the \((\mathrm{IS}_0/N,\mathrm{IS}_1/N)\) plane against the Nash and TWAP benchmarks. In contrast to the previous specifications, many runs now fall in the supra-competitive region defined relative to the Nash benchmark, namely the south-west quadrant in which both players achieve lower implementation shortfall than the Nash point. Since movement in that direction is also movement toward the TWAP benchmark, these outcomes indicate that the learned policies are no longer merely fluctuating around Nash, but are instead shifting systematically toward more cooperative liquidation patterns.
In panel~\textbf{(b)}, for each run we compute the rolling share of training episodes that end in the supra-competitive region, using a \(20\)-episode moving window and the Nash benchmark as the reference point. Relative to the earlier architectures, this share now tends to rise over training in most runs and often remains well above the \(25\%\) benchmark implied by the four-quadrant partition of the plane. 

Taken together, these patterns suggest that access to intra-episode history materially changes the learned behavior. The shift is not limited to a few isolated trajectories or transient excursions; rather, it appears in both the final testing outcomes and the training dynamics themselves. In this sense, incorporating the recent history of prices and own past actions makes sustained supra-competitive outcomes considerably more likely than in the architectures that condition only on the current state.

\begin{figure}[!htbp]
    \centering

    \begin{subfigure}{1\textwidth}
        \centering
        \includegraphics[width=\textwidth]{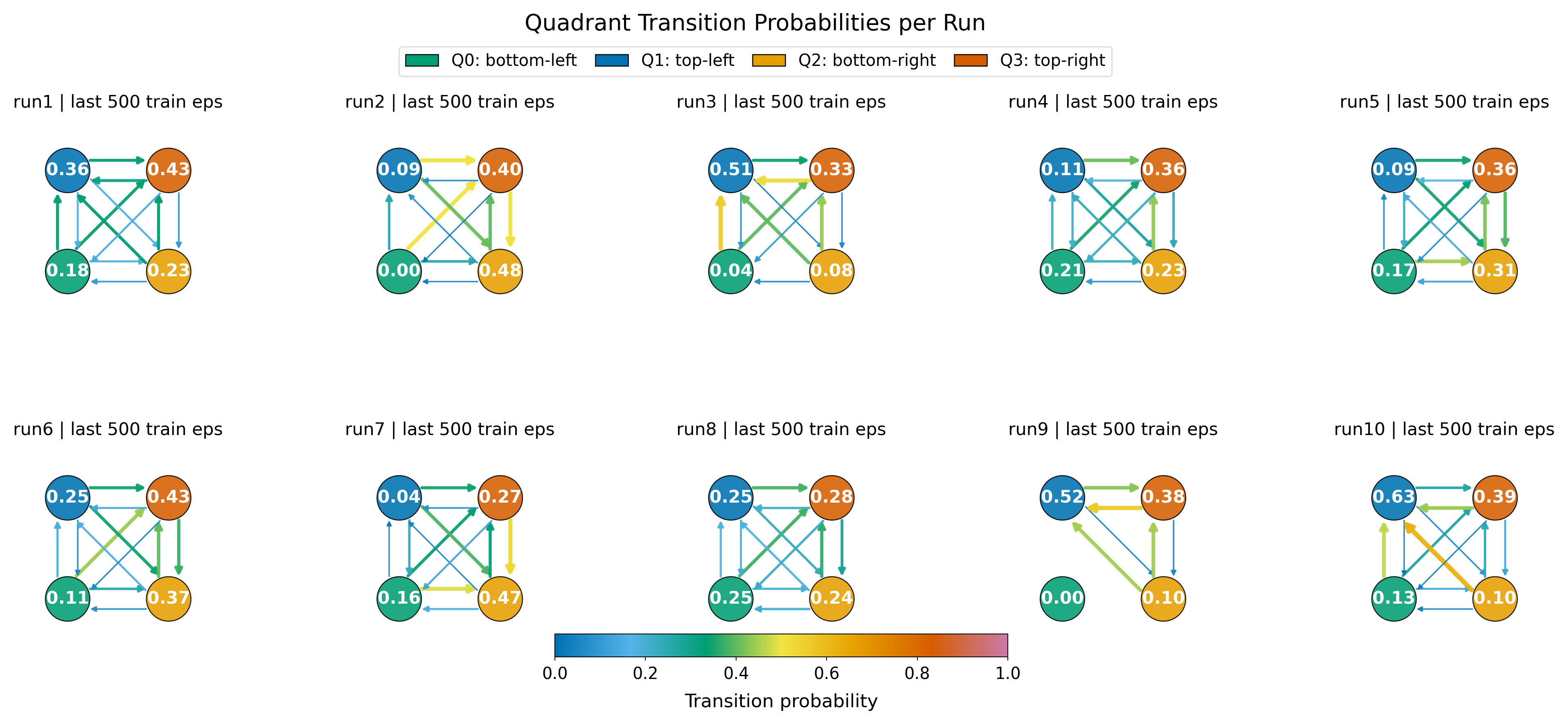}
        \caption{}
        \label{fig:quadrant_transitions_baseline}
    \end{subfigure}

    \vspace{0.8em}

    \begin{subfigure}{1\textwidth}
        \centering
        \includegraphics[width=\textwidth]{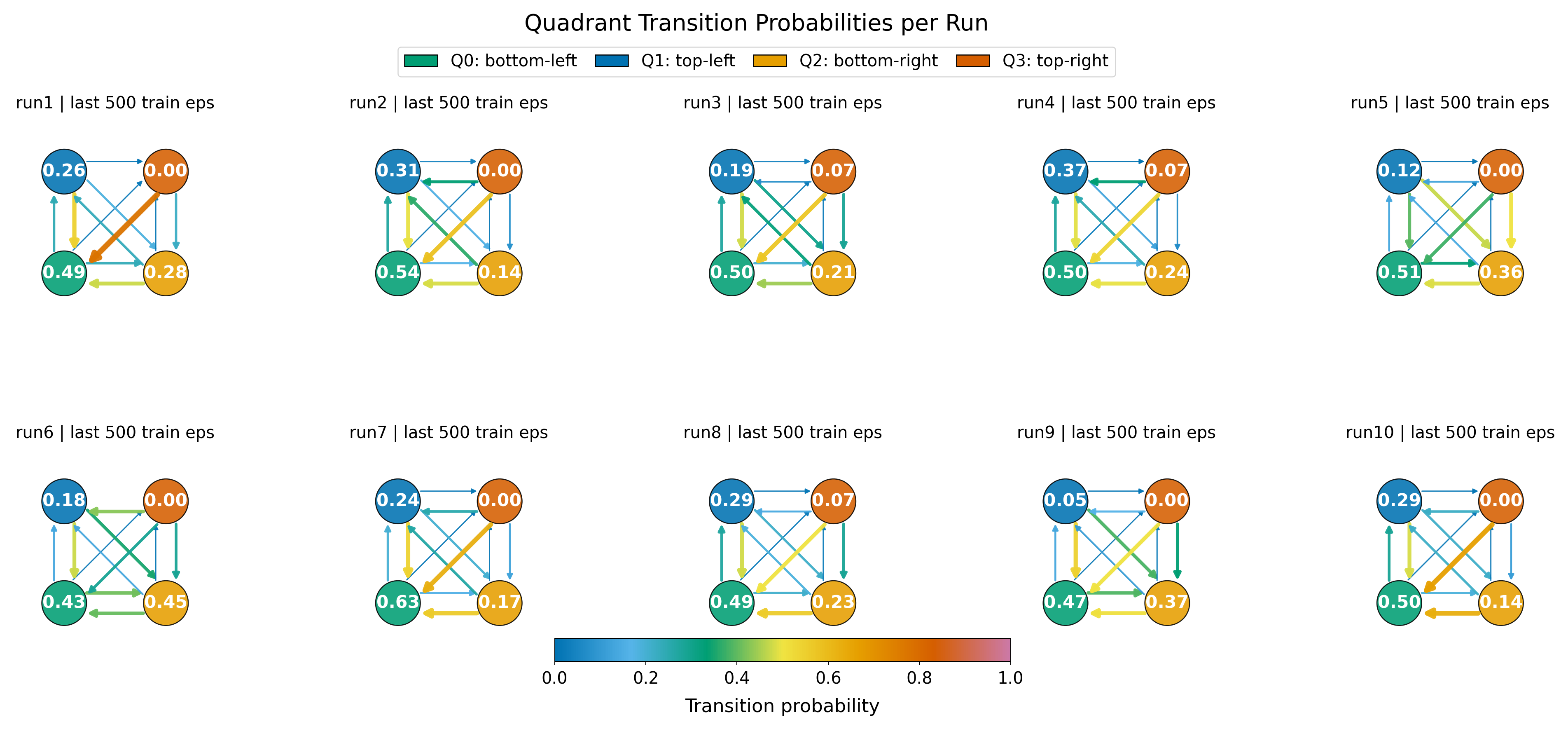}
        \caption{}
        \label{fig:quadrant_transitions_history}
    \end{subfigure}

    \caption{Quadrant occupancies and transition probabilities during the last \(500\) training episodes. \textbf{(a)} Baseline DDQN. \textbf{(b)} History-informed DDQN.}
    \label{fig:quadrant_transitions_comparison}
\end{figure}

A complementary way to assess the dynamics of learning is to examine how training episodes move across quadrants of the \((\mathrm{IS}_1/N,\mathrm{IS}_2/N)\) plane relative to the discrete Nash benchmark during the last \(500\) training episodes. Figure~\ref{fig:quadrant_transitions_comparison} reports, for each run, the empirical quadrant occupancies together with the corresponding transition probabilities. Under the baseline DDQN architecture, the mass remains comparatively dispersed across quadrants and transitions are frequent in several directions, indicating that visits to the supra-competitive region are often temporary and unstable. By contrast, under the history-informed architecture, the dynamics become visibly more concentrated in the bottom-left quadrant, i.e.\ the supra-competitive region in which both players achieve lower implementation shortfall than under the discrete Nash benchmark. At the same time, transitions out of that region appear less prominent, while occupancy of the top-right quadrant becomes very small or vanishes entirely in several runs.

The centroid evidence in Figure~\ref{fig:history_attention_group} shows that once intra-episode history is incorporated into the state representation, the learned outcomes move systematically into the supra-competitive region relative to the discrete Nash benchmark. A natural next question is whether these outcomes are in fact approaching the Pareto-efficient benchmark, or whether they instead stabilize strictly between the competitive and cooperative benchmarks. To answer this, we examine both the average testing inventory paths and the behavior of the testing centroids as the training horizon is increased.

\begin{figure}[!htbp]
    \centering
    \includegraphics[width=\textwidth]{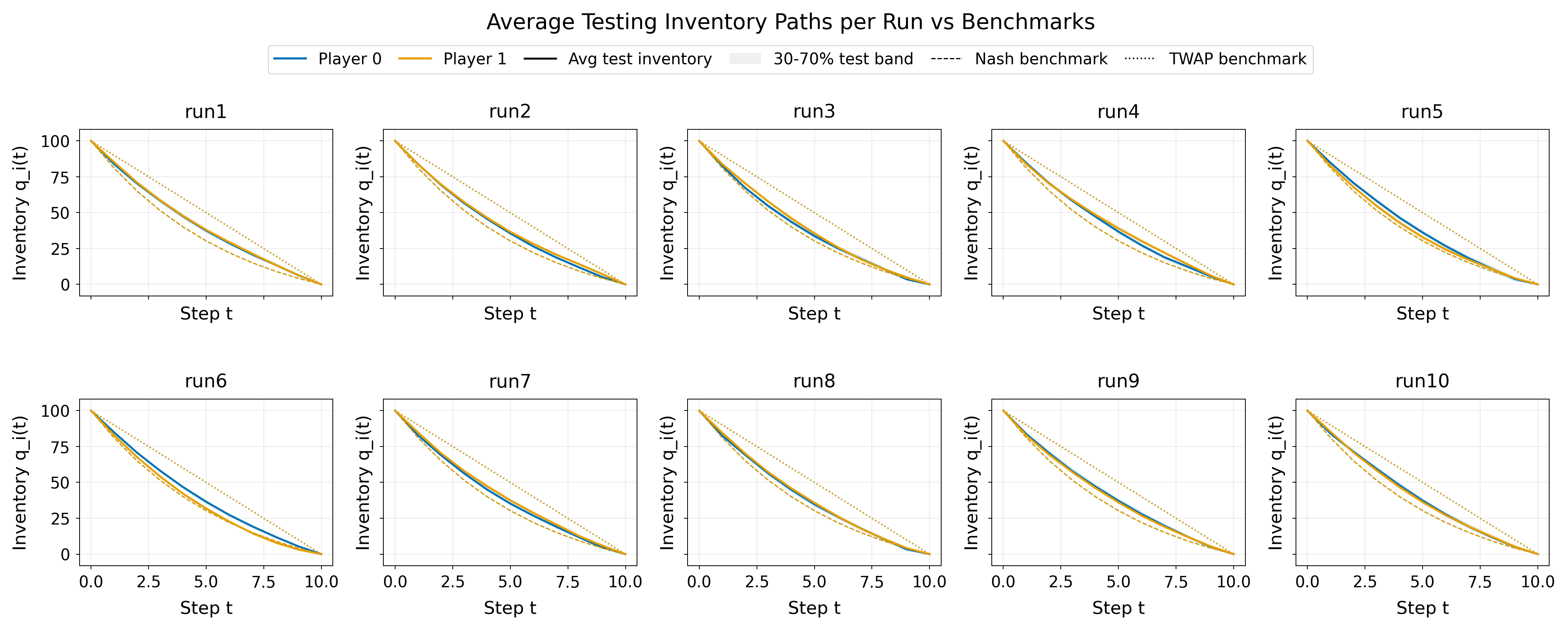}
    \caption{Average testing inventory paths for the history-aware architecture, shown separately for each run. In each panel, the blue and orange curves are the average inventory paths of players 0 and 1 across greedy test episodes in that run, while the dashed and dotted curves are the discrete Nash and discrete TWAP benchmarks. The learned paths lie between the two benchmark paths for most of the episode, indicating liquidation that is slower than Nash but faster than TWAP.}
    \label{fig:history_inventory_paths}
\end{figure}

For run \(r\), player \(i\in\{1,2\}\), and greedy test episode \(m=1,\dots,M\), let \(q_{i,t}^{(r,m)}\) denote the remaining inventory of player \(i\) at step \(t\). We define the average testing inventory path of player \(i\) in run \(r\) by
\[
\bar q_{i,t}^{(r)}
=
\frac{1}{M}\sum_{m=1}^{M} q_{i,t}^{(r,m)},
\qquad t=0,\dots,N.
\]
Figure~\ref{fig:history_inventory_paths} plots these player-specific average testing inventory paths against the discrete Nash and discrete TWAP benchmarks. Across runs, the learned paths lie between the two benchmark paths for most intermediate dates: they are typically above the discrete Nash path and below the discrete TWAP path. Since a higher inventory path corresponds to slower liquidation, this means that the history-aware agents liquidate more slowly than under the discrete Nash benchmark, but still more quickly than under the discrete TWAP benchmark, over most of the episode. This path geometry is consistent with the implementation-shortfall ordering
\begin{equation}
\frac{\mathrm{IS}^{\mathrm{TWAP},\Delta}_i}{N}
<
\frac{\overline{\mathrm{IS}}^{(r)}_i}{N}
<
\frac{\mathrm{IS}^{\text{NE},\Delta}_i}{N},
\qquad i\in\{1,2\},
\label{eq:history_is_ordering}
\end{equation}
where
\[
\overline{\mathrm{IS}}^{(r)}_i
=
\frac{1}{M}\sum_{m=1}^{M}\mathrm{IS}_{i}^{(r,m)}
\]
is the average testing implementation shortfall of player \(i\) in run \(r\). Thus, the learned policies yield lower implementation shortfalls for both players than the competitive benchmark, but higher implementation shortfalls than the Pareto-efficient benchmark. In this precise sense, the outcomes are supra-competitive relative to the discrete Nash benchmark, but not Pareto-efficient.

We next ask whether this gap to the discrete TWAP benchmark disappears if the agents are trained for substantially longer. To investigate this, we repeat the history-aware experiment for training horizons
\[
E \in \{10{,}000,\,20{,}000,\,30{,}000,\,40{,}000,\,50{,}000,\,60{,}000\}.
\]
Across these experiments, we keep fixed the \emph{absolute} number of episodes after which exploration reaches its floor. In the baseline setup, \(\epsilon\) reaches \(\epsilon_{\min}\) after \(4{,}000\) episodes. For each longer-horizon experiment, we therefore adjust the decay schedule so that \(\epsilon\) again first reaches \(\epsilon_{\min}\) at episode \(4{,}000\). Hence the number of exploration-dominated episodes is held constant across experiments, even though the fraction of training until \(\epsilon_{\min}\) is allowed to vary with the total horizon.

For a given horizon \(E\), run \(r\), and player \(i\in\{1,2\}\), let
\[
\overline{\mathrm{IS}}^{(r,E)}_i
=
\frac{1}{M}\sum_{m=1}^{M}\mathrm{IS}^{(r,E,m)}_i
\]
be the average implementation shortfall over the \(M\) greedy test episodes. The corresponding testing centroid in the \((\mathrm{IS}_1/N,\mathrm{IS}_2/N)\) plane is then
\begin{equation}
\bar c^{(r,E)}
=
\left(
\frac{\overline{\mathrm{IS}}^{(r,E)}_1}{N},
\frac{\overline{\mathrm{IS}}^{(r,E)}_2}{N}
\right).
\label{eq:history_centroid_def}
\end{equation}
Let \(c_{N,\Delta}\) and \(c_{\mathrm{TWAP},\Delta}\) denote the discrete Nash and discrete TWAP benchmark points in the same plane. We summarize the position of the learned outcomes by the average Euclidean distances
\begin{equation}
d_N(E)
=
\frac{1}{R}\sum_{r=1}^{R}
\left\|
\bar c^{(r,E)} - c_{N,\Delta}
\right\|_2,
\qquad
d_T(E)
=
\frac{1}{R}\sum_{r=1}^{R}
\left\|
\bar c^{(r,E)} - c_{\mathrm{TWAP},\Delta}
\right\|_2,
\label{eq:history_distance_benchmarks}
\end{equation}
where \(R\) is the number of independent runs.

\begin{figure}[!htbp]
    \centering
    \includegraphics[width=0.88\textwidth]{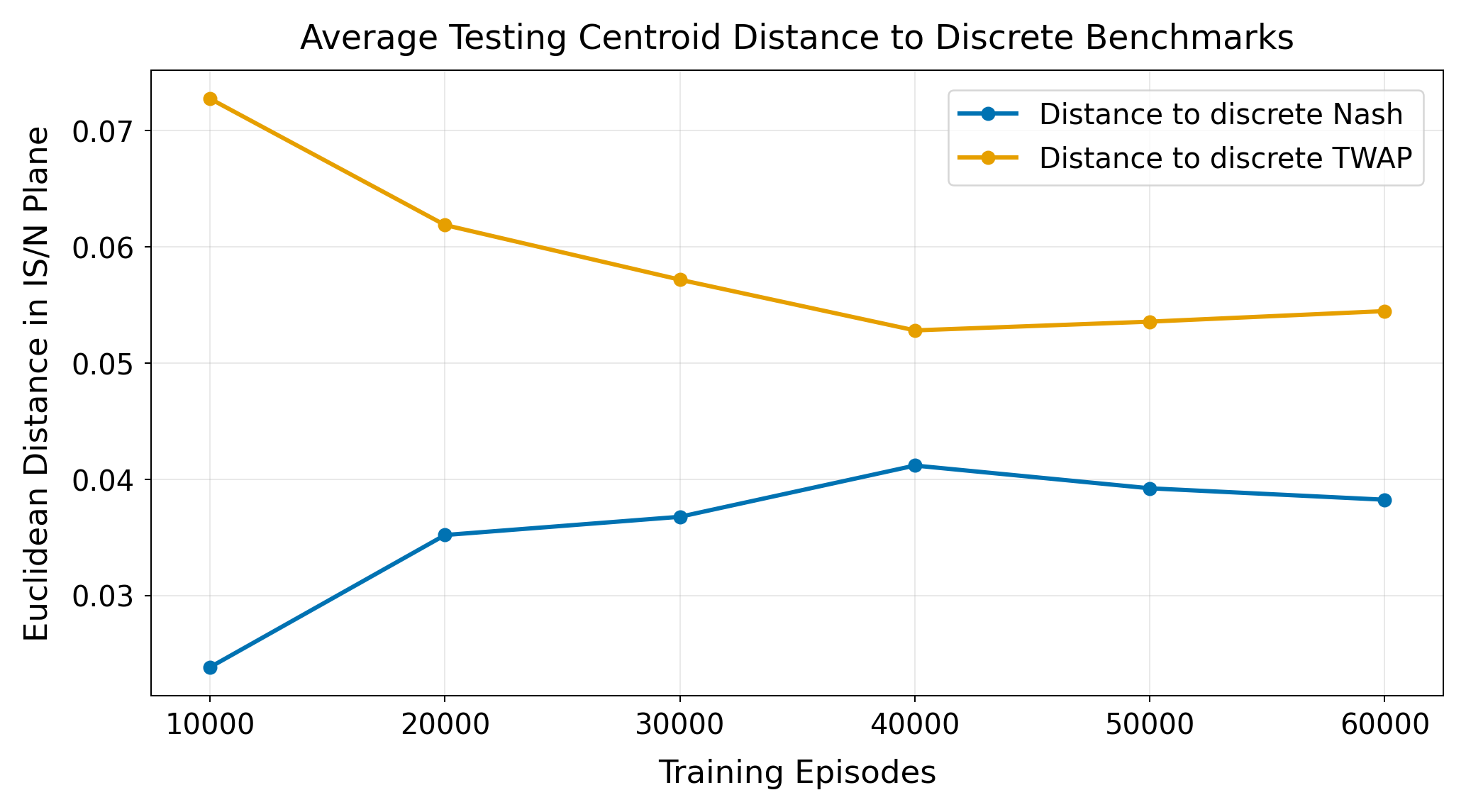}
    \caption{Average Euclidean distance of the final testing centroids to the discrete Nash and discrete TWAP benchmarks as the training horizon increases. For each run, the testing centroid is obtained by averaging \(\mathrm{IS}_1\) and \(\mathrm{IS}_2\) separately over the greedy test episodes and plotting the resulting point in the \((\mathrm{IS}_1/N,\mathrm{IS}_2/N)\) plane. The exploration schedule is adjusted so that \(\epsilon\) reaches \(\epsilon_{\min}\) after the same absolute number of episodes in every experiment. Both distances stabilize away from zero, indicating convergence to an intermediate region rather than to either benchmark exactly.}
    \label{fig:history_distance_vs_horizon}
\end{figure}

Figure~\ref{fig:history_distance_vs_horizon} shows that these distances stabilize at strictly positive values as the training horizon increases. After an initial adjustment phase, the average distance to the discrete Nash benchmark no longer shrinks toward zero, while the average distance to the discrete TWAP benchmark also remains bounded away from zero. Thus, extending training does not drive the learned testing centroids to either benchmark exactly. Instead, the learned outcomes stabilize in an interior region of the \((\mathrm{IS}_1/N,\mathrm{IS}_2/N)\) plane: south-west of the discrete Nash point, but still north-east of the discrete TWAP point.

The shape of the learned inventory paths also helps explain this intermediate behavior. At the beginning of the episode, the available within-episode history is necessarily short, so the sequence branch has only limited information beyond the contemporaneous state. In that phase, the policy remains close to the competitive benchmark, which is why the learned paths start near the discrete Nash trajectory. As the episode unfolds, however, the accumulated history of prices and own past actions becomes increasingly informative, allowing the agent to condition its behavior on a richer description of the realized execution path. Once this additional information becomes available, the learned policy shifts toward a slower and more cooperative liquidation pattern, so that the inventory paths become more TWAP-like over the middle part of the episode. Near the end of the horizon, however, the terminal liquidation constraint \(q_{i,N}=0\) progressively reduces the set of feasible continuations and raises the cost of further delay. The learned paths therefore remain bounded away from the discrete TWAP benchmark and bend back toward full liquidation as maturity approaches.

These additional results sharpen the interpretation of the history-aware specification. Intra-episode history does not push the agents all the way to the Pareto-efficient benchmark. Rather, it induces a stable intermediate regime in which both players achieve lower implementation shortfalls than under the competitive benchmark, while still stopping short of the cooperative optimum.
\section{Conclusions}
\label{sec:conclusions}

In this paper, we revisited the emergence of supra-competitive outcomes in a two-agent Almgren--Chriss execution game and asked which ingredients of the learning problem are responsible for deviations from the competitive benchmark. The central message is that such deviations are not an automatic consequence of multi-agent reinforcement learning. They depend on the design of the execution environment, on the amount of intra-episode feedback available to the agents, and on whether the agents can use that feedback to condition their behavior on the realized evolution of the game.

We first studied what happens when intra-episode feedback is removed. The schedule-learning agents choose complete liquidation trajectories at the beginning of the episode and cannot react to within-episode price movements, realized inventories, or traces of the opponent's behavior. This matters because much of the broader literature links supra-competitive outcomes to feedback channels, including implicit signalling, reward--punishment dynamics and other relevant behaviours. By removing these channels, the schedule-learning experiments isolate what can arise purely from ex-ante schedule selection.

Under aggregate temporary impact, removing intra-episode feedback leads both schedule-learning procedures to recover the corresponding Nash benchmark. This shows that, in the simultaneous aggregate-impact environment, learning does not by itself move agents persistently away from the competitive schedule when the agents cannot react within the episode. In that case, the learned schedules remain concentrated around the aggregate-temporary-impact Nash path.

The own-temporary-impact schedule-learning experiment is more subtle. From a strict comparison with the own-temporary-impact Nash benchmark, the learned outcomes appear supra-competitive: both agents achieve lower implementation shortfalls than the own-temporary-impact Nash point. However, the path-level evidence changes the interpretation. The learned schedules do not move toward TWAP, nor toward an arbitrary cooperative trajectory. Instead, they lie very close to the aggregate-temporary-impact Nash path. The auxiliary experiment in which one player is fixed at the aggregate-temp Nash schedule reinforces this reading: the learning player responds by remaining close to that same schedule, rather than exploiting the fixed opponent.

This suggests that the apparent supra-competitive outcome under own temporary impact is partly a limited-information phenomenon. The own- and aggregate-temporary-impact environments differ in the execution price and therefore in realized payoffs, but they share the same inventory and midprice transition law. When agents commit to full schedules ex ante and receive no intra-episode feedback, the dominant dynamic signal remains the aggregate permanent-impact channel. The own-temp convention affects the payoff accounting, but it does not generate a distinct state-transition signal that naturally guides the learner toward the own-temp Nash path. Thus, from the strict own-temp Nash point of view, the outcome is supra-competitive; at the level of learned inventory paths, it is better understood as convergence toward an aggregate-Nash-like effective benchmark under limited information flow.

We then allowed agents to receive intra-episode feedback through a state-dependent DDQN architecture. Relative to \citet{lillo2024deviations}, we made three design choices to isolate the mechanism more cleanly: agents act simultaneously within each timestep, exploration is not centered around TWAP, and the baseline environment uses aggregate temporary impact. These choices remove within-slice information leakage, avoid building the cooperative schedule into the exploration process, and align the simulator with the aggregate-impact benchmark. In this setting, baseline DDQN agents do not produce persistent supra-competitive outcomes. Testing centroids remain close to the aggregate-temp Nash benchmark, and the rolling share of training episodes in the supra-competitive region does not stabilize at high levels.

The integrated-gradient diagnostics help explain this result. Although the agents observe the current price, the learned value function assigns little effective importance to the price coordinate after the initial training phase. The Q-function is driven primarily by remaining inventory and time. At the same time, the agents do not collapse to TWAP. If they were ignoring the strategic nature of the environment altogether, one would expect behavior closer to the single-agent and Pareto-efficient TWAP schedule. Instead, the learned policies remain near the Nash benchmark, suggesting that the agents do recognize the competitive structure of the execution game, but do not use contemporaneous price as a coordination channel.

Strengthening the price channel changes the attribution pattern but not the economic conclusion. Architectures that make the current price more salient assign greater importance to the price input, yet they still do not generate a systematic and persistent movement toward the supra-competitive region. This indicates that the relevant missing ingredient is not simply price observation or price sensitivity. What matters is not only the current price level, but the history of how that price was reached.

The qualitative picture changes once agents can condition on intra-episode history. When the state representation is augmented with recent prices and own past actions, supra-competitive outcomes become more frequent and more stable. The learned policies achieve lower implementation shortfalls for both players than the competitive Nash benchmark, while remaining bounded away from the Pareto-efficient TWAP benchmark. In this sense, the learned behavior is genuinely supra-competitive but not fully cooperative. The inventory paths clarify the mechanism: early in the episode, when little history is available, behavior remains close to the competitive benchmark; as prices and actions accumulate, the agents shift toward slower, more TWAP-like liquidation; near maturity, the terminal liquidation constraint forces the policies back toward full liquidation. Longer training does not eliminate this gap. The learned outcomes stabilize between Nash and TWAP rather than converging exactly to either benchmark.

Taken together, the results show that supra-competitive outcomes in this execution game require more than multi-agent learning and more than observation of the current price. They arise when agents can process the realized evolution of the episode and condition their actions on accumulated history. This places the mechanism closer to feedback, memory, and state-contingent interaction than to price observation alone. The paper therefore contributes to the study of algorithmic interaction in market microstructure by showing that both the information set and the architecture of learning agents can qualitatively change the economic outcome of an execution game.

More broadly, our findings emphasize the need to interpret learned trading behavior jointly with the environment design and information structure under study. Removing intra-episode feedback can make agents behave as if they face a different effective benchmark; allowing state-dependent feedback alone does not necessarily generate persistent supra-competitive outcomes; and history-aware policies can support outcomes that are neither Nash nor fully cooperative. Understanding these distinctions is essential for determining when learned execution behavior should be viewed as competitive, supra-competitive, or an artifact of the benchmark and information structure used for comparison.

\appendix
\renewcommand{\thesection}{Appendix~\Alph{section}}
\section{Model-based ex-ante schedule learning}
\label{app:deterministic_surrogate_schedule_learning}

Here we describe the model-based ex-ante schedule learner used as a robustness check for the experiments in Section~\ref{sec:open_loop_rl}. As in the model-free procedure, each agent chooses a complete liquidation schedule before the episode begins and then executes that schedule without conditioning on the realized intra-episode state trajectory.

The learner uses a time-only neural network to parameterize the trading schedule. In this implementation, the first \(N-1\) actions are parameterized around a TWAP initialization:
\[
\widetilde U_t^{(k)}
=
\frac{q_0^{(k)}}{N}
+
\phi_{\theta^{(k)}}(t/N)
+
b_t^{(k)},
\qquad
t=0,\dots,N-2.
\]
The final action is then chosen to enforce exact liquidation:
\begin{equation}
U_{N-1}^{(k)}
=
q_0^{(k)}
-
\sum_{t=0}^{N-2}
\widetilde U_t^{(k)}.
\label{eq:app_ol_mb_terminal}
\end{equation}
Hence
\[
\sum_{t=0}^{N-1}U_t^{(k)}=q_0^{(k)}.
\]

The model-based learner differs from the model-free learner in how local dynamics information is obtained. Instead of estimating the effect of schedule perturbations entirely from simulator rollouts, it uses the deterministic transition approximation
\begin{equation}
q_{t+1}^{(1)}
=
q_t^{(1)}
-
u_t^{(1)},
\qquad
q_{t+1}^{(2)}
=
q_t^{(2)}
-
u_t^{(2)},
\qquad
S_{t+1}
=
S_t
-
\kappa
\bigl(u_t^{(1)}+u_t^{(2)}\bigr).
\label{eq:app_ol_mb_dyn}
\end{equation}
For agent \(k\), the private state is
\[
x_t^{(k)}
=
\begin{bmatrix}
q_t^{(k)}\\
S_t
\end{bmatrix}.
\]
Given the opponent's current schedule, the private dynamics can be written as
\begin{equation}
x_{t+1}^{(k)}
=
x_t^{(k)}
+
\begin{bmatrix}
-u_t^{(k)}\\
-\kappa\bigl(u_t^{(k)}+u_t^{(-k)}\bigr)
\end{bmatrix}.
\label{eq:app_ol_mb_private_dyn}
\end{equation}
Therefore the local derivatives of the private transition with respect to the private state and own action are
\begin{equation}
A_t^{(k)}
=
I_2,
\qquad
B_t^{(k)}
=
\begin{bmatrix}
-1\\
-\kappa
\end{bmatrix}.
\label{eq:app_ol_mb_jacobians}
\end{equation}

The model-based update then uses these local derivatives inside the same schedule-improvement logic as the model-free procedure. The key distinction is therefore the source of local dynamics information. In both cases, the learned object is a complete ex-ante liquidation schedule rather than a state-dependent feedback rule.

Across the experiments reported in Section~\ref{sec:open_loop_rl}, the model-based learner produced qualitatively similar results to the model-free ex-ante learner.

\bibliographystyle{apalike}
\bibliography{refs}

\end{document}